%% file: main.tex
\documentclass[a4paper,twocolumn,11pt,accepted=2026-03-09]{quantumarticle}
\pdfoutput=1
\usepackage[utf8]{inputenc}
\usepackage[english]{babel}
\usepackage[T1]{fontenc}
\usepackage{amsmath}
\usepackage{hyperref}
\usepackage{amsfonts}
\usepackage{bm}

\usepackage{tikz}
\usepackage{lipsum}

\usepackage{xcolor}

\newtheorem{theorem}{Theorem}

\usepackage[sort&compress,numbers,square,comma]{natbib}

\begin{document}

\title{Quantum correlations in the steady state of light-emitter ensembles from perturbation theory}

\author{Dolf Huybrechts}
\email{dolf.huyb@gmail.com}
\affiliation{Univ Lyon, Ens de Lyon, CNRS, Laboratoire de Physique, F-69342 Lyon, France}
\affiliation{Advanced Concepts Team, European Space Agency,
European Space Research and Technology Centre, 2201 AZ Noordwijk, The Netherlands}
\orcid{0000-0002-5821-3493}
\author{Tommaso Roscilde}
\email{tommaso.roscilde@ens-lyon.fr}
\affiliation{Univ Lyon, Ens de Lyon, CNRS, Laboratoire de Physique, F-69342 Lyon, France}

\maketitle

\begin{abstract}
 The coupling of a quantum system to an environment leads generally to decoherence, and it is detrimental to quantum correlations within the system itself. Yet some forms of quantum correlations can be robust to the presence of an environment  -- or may even be stabilized by it. Predicting (let alone understanding) them remains arduous, given that the steady state of an open quantum system can be very different from an equilibrium thermodynamic state; and its reconstruction requires generically the numerical solution of the Lindblad equation, which is extremely costly for numerics. Here we focus on the highly relevant situation of ensembles of light emitters undergoing spontaneous decay; and we show that, whenever their Hamiltonian is perturbed away from a U(1) symmetric form, steady-state quantum correlations can be reconstructed via pure-state perturbation theory. Our main result is that in systems of light emitters subject to single-emitter or two-emitter driving, the steady state perturbed away from the U(1) limit generically exhibits spin squeezing; and it has minimal uncertainty for the collective-spin components, revealing that squeezing represents the optimal resource for entanglement-assisted metrology using this state.     
\end{abstract}

\section{Introduction}

 Quantum correlations \cite{Frerot2023}, and in particular many-body entanglement \cite{Horodeckietal2009}, are the central resource of quantum many-body devices processing quantum information or simulating complex interacting Hamiltonians \cite{Georgescuetal2014}. The ability to engineer arbitrary entangled quantum states of \emph{e.g} a system of $N$ qubits amounts in principle to controlling an exponentially large (in $N$) amount of information -- as opposed to linear in $N$ for classical information. 
 Nonetheless quantum information of $N$ qubits is as astronomically large as it is fragile to the presence of an environment. Since information is stored in the relative phases and amplitudes of many-body superposition states, it is extremely sensitive to decoherence and dissipation, generically reducing entangled states to statistical mixtures of states. To counter this effect, the most obvious strategy is to decouple qubits from their environment as much as possible. Yet a more specialized strategy to preserve entanglement in open quantum systems consists in finding forms of entanglement which are \emph{robust} to the presence of an environment, or that are even stabilized by it. Such robustness to, or assistance from the environment can come from distinct mechanisms. One relevant example is collective dissipation in driven-dissipative systems, in which the degrees of freedom are driven individually, but the elementary dissipation processes can induce entanglement in the system \cite{VerstraeteNATPH2009}. Through an appropriate engineering of the coupling of the system to the environment, one can obtain steady state entanglement and complex non-equilibrium phases of matter \cite{VerstraeteNATPH2009, DiehlNATPH2008, WeimerNPhys2010, Mueller_2012}. Another important example is the competition between dissipation (individual or collective) and entangling Hamiltonian dynamics \cite{LeePRL13, JinPRX16, Verstraelenetal2023, OverbeckPRA17}. As a result, the stationary state of the dissipative quantum system can feature entanglement going beyond the paradigm of unitary evolutions, or beyond that of ground-state physics for interacting Hamiltonians. 
 
 Understanding the kind of entangled states which can emerge as stationary states of driven-dissipative many-body systems is a formidable theoretical challenge. Indeed reconstructing the stationary state of generic dissipative systems is numerically prohibitive already in systems of $\sim O(10)$ elementary degrees of freedom (such as qubit ensembles). And approximation schemes to tackle the stationary state are often limited in the amount of entanglement that they can describe, either within the system itself (such as in mean-field approaches \cite{JinPRX16}); or of the system with the environment (such as in approaches based on reduced Hilbert spaces \cite{FinazziPRL15}). A significant effort has been devoted in recent years to the development of methods that can accurately describe the properties of open quantum systems \cite{JinPRX16, FinazziPRL15, WeimerPRL2015, Sieberer_2016, ShammahPRA98, VicentiniPRL19, NagyPRL19, HartmannPRL19, Weimeretal2021, DeuarPRXQ21, mink2022hybrid, Verstraelenetal2023}. However, the understanding of their correlation properties remains very challenging, and difficult to guess without a dedicated microscopic calculation.

 In this work we offer a framework to understand and predict the correlation properties of stationary states of an important class of dissipative quantum systems, i.e. ensembles of quantum emitters coupled to the vacuum of the electromagnetic field \cite{ReitzPRXQuantum2022}. If the Hamiltonian governing the unitary dynamics of the system has rotation (U(1)) symmetry along the quantization axis of the emitters, then the steady state of the system is trivial, and it corresponds to all the emitters in their ground state. We show that a U(1)-symmetry-breaking perturbation in the Hamiltonian, if treated within pure-state perturbation theory, induces quantum correlations in the steady state. This happens at first order in the perturbation, if the perturbation drives pairs of emitters; or at second order, if the perturbation drives individual emitters independently. In particular, the perturbed states possess a very clear form of quantum correlations, i.e. \emph{spin squeezing}, and they have minimal uncertainty for the collective-spin operators, so that squeezing is their optimal metrological resource \cite{PezzeRMP2018}.  As we shall detail in the paper, pure-state perturbation theory is fully justified when quantum correlations in the steady state build up faster with the perturbation than entropy does -- as observed in all the examples we considered here.
 The perturbative calculations can be fully worked out in the case of collective-spin models with individual or collective emission; they allow one to understand the steady state as a perturbed, arbitrarily excited eigenstate of the system's Hamiltonian;  and they can be tested successfully against exact results. 
 The insight gained from these reference models serves as a very useful template for the understanding of more complex situations of models with spatially structured interactions and/or spatially structured collective emission, of direct relevance to current experiments on collective light-matter interactions \cite{Cong16,Ferioli21}. 
 
 Our paper is structured as follows: Sec.~\ref{sec:perttheory} discusses the general setting of pure-state perturbation theory for the steady state of the dissipative dynamics; Sec. ~\ref{sec:emitters} discusses the models of dissipation of interest to this work; Sec.~\ref{sec:collective_spin} introduces the collective-spin properties used to characterize quantum correlations; Sec~\ref{sec:two_emitter} applies the perturbative approach to the case of models with two-emitter driving,  formulating the condition under which squeezing is expected in the steady state in the form of a "squeezing theorem" for dissipative steady states; Sec.~\ref{sec:applications} reconstructs the squeezing properties for the collective XYZ model and transverse-field Ising model with single-emitter dissipation; Sec.~\ref{sec:drivenDicke} discusses instead models with single-emitter drive and collective dissipation, with special focus on the driven Dicke model. A discussion of the experimental implications and conclusions are offered in Secs.~\ref{sec:discussion} and ~\ref{sec:conclusions}.

 \begin{figure*}[t]
  \centering
  \includegraphics[width=\textwidth]{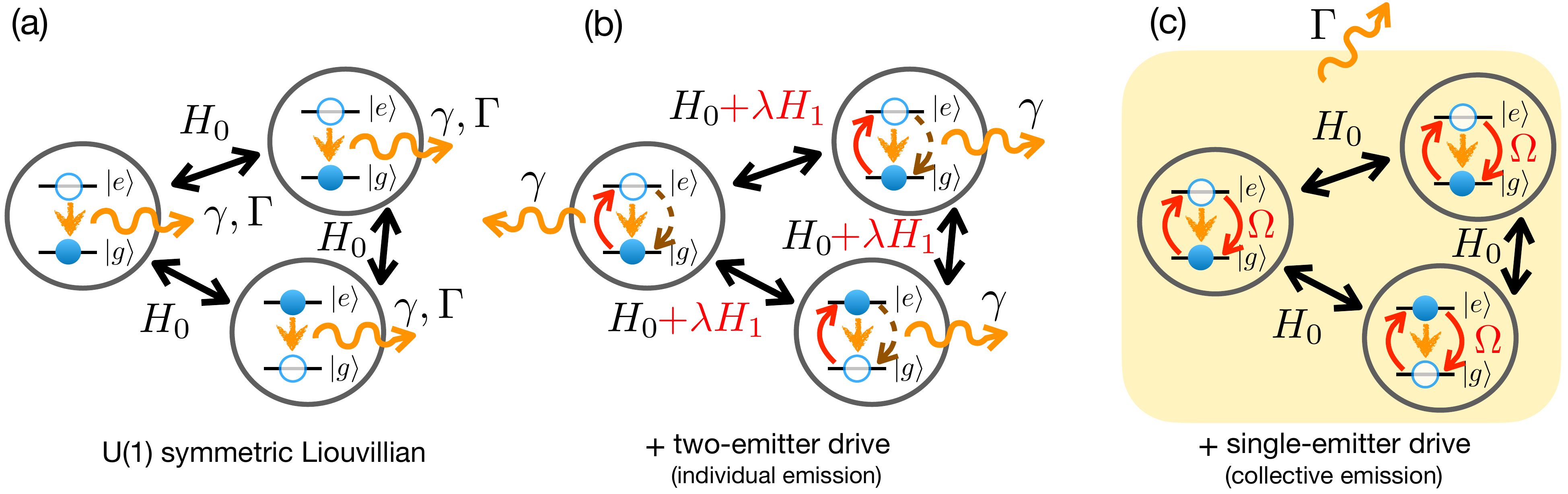}
  \caption{Sketch of the physical situations of interest to this work: (a) the unperturbed reference is a system of light emitters interacting via a U(1)-symmetric Hamiltonian, and emitting photons in the environment; (b) a Hamiltonian perturbation $\lambda H_1$ associated with a two-emitter drive, re-exciting (or de-exciting) the emitters in pairs, can break the U(1) symmetry; (c) a single-emitter (Rabi) drive with strength $\Omega$.}
  \label{fig:sketch}
\end{figure*}

\section{Pure-state perturbation theory for the steady state of the Lindblad equation}\label{sec:perttheory}
 
 \subsection{General setting}
 
The focus of our work are open quantum systems that interact with a Markovian environment, and can be described by a Lindblad master equation \cite{BreuerBookOpen}
\begin{equation}\label{eq:lindblad}
    \partial_t{\rho} = {\cal L}[\rho] = -i\left[{H}, {\rho}\right] + \sum_j \mathcal{D}\left(\sqrt{\gamma_j} ~{L}_j\right)[\rho],
\end{equation}
where ${\cal L}[.]$ denotes the Liouvillian superoperator,  ${H}$ is the system's Hamiltonian, ${\rho}$ the system's density matrix, and $\mathcal{D}\left(\sqrt{\gamma_j} ~ {L}_j\right)$ represents the dissipative process associated with the jump operators $ \sqrt{\gamma_j}  ~{L}_j$, and it reads
\begin{equation}
    \mathcal{D}\left(\sqrt{\gamma_j} ~{L}_j\right)[\rho] = \frac{\gamma_j}{2}\left(2{L}_j{\rho}{L}_j^\dagger - \left\{{L}_j^\dagger{L}_j,{\rho} \right\}\right).
\end{equation}

Our goal is the search of the steady state $\rho_{ss}$ such that ${\cal L}[\rho_{\rm ss}]=0$. The general setting for perturbation theory of Markovian open quantum systems
\cite{Benatti_JPA2011, Li14, LiPRX16,GomezRiP18, LenarcicPRB18, KatoBOOK, Kessler2012}  assumes that the Liouvillian can be written as ${\cal L} = {\cal L}_0 + \lambda {\cal L}_1$, where ${\cal L}_0[.] = -i[H_0,(.)] + \sum_j \mathcal{D}_0\left(\sqrt{\gamma_j} ~{L}_j\right)[.]$ is an unperturbed Liouvillian for which the steady state ($\rho_{\rm ss} = \rho_0$) is known; while $\lambda {\cal L}_1$ is a perturbation, parametrized by $\lambda \in \mathbb R$. Since our focus is on elementary mechanisms inducing quantum correlations in the steady state, we shall specialize our interest to the case in which
\begin{enumerate}
\item the unperturbed steady state is a unique, \emph{pure  state}, $\rho_0 = |\phi_0 \rangle \langle \phi_0|$, corresponding to an eigenstate of $H_0$, and such that, for all $j$, $L_j |\phi_0\rangle=0$;
\item the perturbation is purely Hamiltonian, and the perturbed Hamiltonian reads $H = H_0 + \lambda H_1$.  
\end{enumerate}

The relevance of the above assumptions to experiments is rather clear. The steady state of systems of $N$ degrees of freedom coupled to a dissipative environment is typically one in which each degree of freedom loses all its energy to the environment, and falls to its single-body ground state, so that $\sum_j \mathcal{D}_0\left(\sqrt{\gamma_j} ~{L}_j\right)[\rho_0] = 0$; and at the same time this state is also a (generically excited) eigenstate of the many-body Hamiltonian $H_0$,  ${H}_0 |\phi_0\rangle = E_0 |\phi_0\rangle$ so that $[H_0,  \rho_0] = 0$. 
This is the case \emph{e.g.} of light emitters (real or artificial atoms) which are not driven, and which therefore decay to their single-body ground state after having emitted. 
A most natural way of perturbing this steady state in an experiment is to perturb the Hamiltonian with a driving term $\lambda H_1$ which alters its symmetry, such that $[H_0 + \lambda H_1,  \rho_0] \neq 0$. 

We want to stress that the perturbation $\lambda H_1$, although involving the Hamiltonian only, should be still viewed as a perturbation to the Liouvillian super-operator, i.e. as 
$\lambda {\cal L}_1[.] = i \lambda [H_1,(.)]$. And its smallness should not be measured in comparison with the Hamiltonian term of the Liouvillian only (which may be vanishing, as we will see in the examples offered in this work); but also in comparison with the dissipative part.

 \subsection{Pure-state perturbation theory}\label{sec:pure}

The perturbed steady state can be written in general as a power series of the perturbation
\begin{equation}
\rho_{\rm ss} = \rho_0 + \lambda \rho_1 + \lambda^2 \rho_2 + ...
\label{e.expansion} 
\end{equation}
and this state will be generically a mixed state. Yet, since $\rho_0$ is pure, we can show  that the corrections $\lambda \rho_1$,  $\lambda^2 \rho_2$, etc. weakly alter its purity, so that the perturbed state can in fact be searched for in the form of a \emph{pure} state as well.
In general the purity of the steady state reads 
\begin{equation}
{\rm Tr}(\rho_{\rm ss}^2) = {\rm Tr}\rho_0 + \lambda {\rm Tr} (\rho_0 \rho_1 + \rho_1 \rho_0) + O(\lambda^2)~.
\end{equation}
Yet one can formally prove that $\rho_0$ and $\rho_1$ are orthogonal to each other (see App.~\ref{sec:proof}), so that the linear term in $\lambda$ vanishes in the above expression. A simple explanation for this aspect is that ${\rm Tr}(\rho_{\rm ss}^2)$ is maximal (and equal to 1) for $\lambda = 0$, and it can be assumed to be a continuous and differentiable function of $\lambda$ around its maximum. This implies therefore that ${\rm Tr}(\rho_{\rm ss}^2)  = 1 + O(\lambda^2)$. Further numerical evidence for this argument is also offered in App. \ref{sec:proof}. 

Within the pure assumption, we write $\rho_{\rm ss}$ as $|\phi_0'\rangle \langle \phi_0' | $ with 
\begin{equation}
|\phi_0'\rangle = |\phi_0\rangle + \lambda | \psi_1 \rangle +  \lambda^2 | \psi_2 \rangle + ...
\label{e.expansion_pure}
\end{equation}
so that 
\begin{align}
\rho_{ss} & \approx  ~ \rho_0 + \lambda \left ( |\phi_0\rangle \langle \psi_1 | + 
|\psi_1 \rangle \langle \phi_0 | \right ) \nonumber \\
&+ \lambda^2\left(\vert\psi_1\rangle\langle\psi_1\vert + \vert\psi_2 \rangle\langle\phi_0\vert + \vert\phi_0\rangle\langle\psi_2 \vert \right) + O(\lambda^3)~.
\end{align}

Here we shall provide the main results of pure-state perturbation theory relevant for the study provided below.

Injecting the expansion Eq.~\eqref{e.expansion} into the Lindblad equation, imposing the condition of stationarity, and equating terms of same order in $\lambda$, one obtains the following conditions on the first- and second-order corrections to the steady state:
\begin{align}
\label{eq:firstorderpert} 
\dot \rho_1 & = -i[H_0,\rho_1] - i[H_1,\rho_0] \\
&+ \sum_j \gamma_j \left [  L_j \rho_1 L_j^\dagger - \frac{1}{2}\{ L^\dagger_j L_j, \rho_1 \} \right ]= 0, \nonumber
\end{align}
\begin{align}
\label{eq:secondorderpert} 
\dot\rho_2 & = -i[H_0,\rho_2] -i[H_1,\rho_1]  \\
&+   \sum_j \gamma_j \left [  L_j \rho_2 L_j^\dagger - \frac{1}{2}\{ L^\dagger_j L_j, \rho_2 \} \right ] = 0. \nonumber
\end{align}

Given the Hamiltonian eigenbasis $H_0 |\phi_n\rangle = E_n |\phi_n\rangle$, and assuming the unperturbed steady state $|\phi_0\rangle$ to be a \emph{non-degenerate} eigenstate of $H_0$, we can decompose the pure-state perturbations $|\psi_1\rangle$ and $|\psi_2 \rangle$ in Eq.~\eqref{e.expansion_pure} on the $\{ |\phi_n\rangle\}$ basis as
 \begin{align}
    \vert\psi_1\rangle & = \sum_{n\neq 0}c_n\vert\phi_n\rangle \nonumber \\
    \vert\psi_2\rangle & = \left ( 1-\frac{1}{2} \langle \psi_1|\psi_1\rangle \right ) \vert\phi_0\rangle + \sum_{n\neq 0} d_n\vert\phi_n\rangle~.
    \label{e.decomposition}
\end{align}
As in ordinary perturbation theory for Hamiltonian eigenstates \cite{Landau_BOOK_Quantum}, the first-order correction can be taken to be orthogonal to the unperturbed state, while the second-order one needs to have overlap with the unperturbed state in order to enforce normalization of the perturbed state. 

Imposing the conditions of stationarity, Eqs.~\eqref{eq:firstorderpert}-\eqref{eq:secondorderpert}, on the states $|\psi_1\rangle$ and $|\psi_2\rangle$, and projecting them onto the Hamiltonian eigenstates $\langle \phi_n |$ from the left and $|\phi_0\rangle$ from the right, leads to the following coupled equations for the $c_n$ and $d_n$ coefficients:
\begin{align}\label{eq:cn}
& (E_n-E_0) c_n  + \langle \phi_n | H_1 | \phi_0 \rangle \\
&- \frac{i}{2} \sum_m c_m \langle \phi_n |  \sum_j \gamma_j L^\dagger_j L_j |\phi_m \rangle = 0,~ \nonumber
\end{align}
\begin{align}\label{eq:dn}
    &\left(E_n - E_0 \right) d_n
    + \langle\phi_n\vert H_1\vert\psi_1\rangle  - c_n\langle\phi_0\vert H_1\vert\phi_0\rangle  \nonumber \\
    &+ i\sum_j  \gamma_j \langle\phi_n\vert L_j \vert\psi_1\rangle\langle\psi_1\vert L_j^\dagger \vert\phi_0\rangle \nonumber \\
    &-\frac{i}{2} \sum_j \sum_m \gamma_j \langle\phi_n\vert L_j^\dagger L_j\vert\phi_m \rangle d_m  = 0
\end{align}
where in the second equation we have kept the $|\psi_1\rangle$ symbol for the sake of brevity. 

The above equations clearly provide the formal solution to the perturbation problem if the matrix elements of the Hamiltonian perturbation $H_1$ and of the jump operators $L_j, L_j^\dagger$ on the basis of the unperturbed Hamiltonian eigenstates are known. We would like to stress that this aspect is made possible because we are restricting our attention to the steady state of the Lindblad dynamics, and we are taking it as a pure state. Previous works \cite{Benatti_JPA2011, Li14, LiPRX16} have rather tackled the more ambitious problem of reconstructing the full perturbed Liouvillian spectrum and the corresponding generalized eigenvectors. This problem is significantly more involved, as the equations defining the perturbation to the generalized eigenvectors require the pseudo-inversion of non-Hermitian super-operators.

\subsection{Special case: $\sum_j \gamma_j L^\dagger_j L_j$ commutes with the unperturbed Hamiltonian}
\label{s.special}

A special case -- relevant to all the examples that we shall provide in the following -- is the one in which the jump operators satisfy the following condition
\begin{equation}
[\sum_j  \gamma_j L^\dagger_j L_j, H_0] = 0~.
\label{e.commutation}
\end{equation}

For this choice of jump operators, $\sum_j  \gamma_j L^\dagger_j L_j$ is diagonal on the eigenbasis of $H_0$, simplifying significantly the equations Eqs.~\eqref{eq:cn}-\eqref{eq:dn} above. Indeed this allows for an explicit expression of the first- and second-order corrections to the pure stationary state:
\begin{align}
|\psi_1\rangle = - \sum_{n\neq 0} \frac{\langle \phi_n | H_1 | \phi_0 \rangle }{\langle H_0^{\rm (NH)} \rangle_n - E_0}~ |\phi_n\rangle
\label{e.psi1}
\end{align}
\begin{align}
 & |\psi_2\rangle =  \nonumber \\
& \sum_{n\neq 0,m \neq 0} \frac{\langle \phi_n | H_1 | \phi_m \rangle  \langle \phi_m | H_1 | \phi_0 \rangle}{(\langle  {H}_0^{\rm (NH)} \rangle_n - E_0)(\langle {H}_0^{\rm (NH)}  \rangle_m - E_0)} ~ |\phi_n\rangle \nonumber \\
& - \sum_{n \neq 0} \frac{ \langle \phi_0 | H_1 |\phi_0 \rangle \langle \phi_n | H_1 |\phi_0 \rangle }{ (\langle {H}_0^{\rm (NH)} \rangle_n - E_0)^2} |\phi_n \rangle \nonumber \\
& -\frac{1}{2} \sum_{n \neq 0} \frac{  |\langle \phi_n | H_1 |\phi_0 \rangle|^2}{\left | \langle {H}_0^{\rm (NH)}  \rangle_n  - E_0 \right | ^2} ~|\phi_0\rangle~ \nonumber \\
& - i \sum_j \gamma_j \sum_{n \neq 0} \frac{\langle \phi_n | L_j |\psi_1 \rangle \langle \psi_1 | L_j^\dagger |\phi_0\rangle}{ \langle {H}_0^{\rm (NH)} \rangle_n - E_0} ~ |\phi_n\rangle 
\label{e.psi2}
\end{align}

Here we have introduced the unperturbed, non-Hermitian (NH)  Hamiltonian
\begin{equation}
    {H}_0^{\rm (NH)} = {H}_0 - i\sum_j\frac{\gamma_j}{2} {L}_j^\dagger {L}_j,
\end{equation}
such that $\langle {H}_0^{\rm (NH)} \rangle_n = \langle \phi_n | {H}_0^{\rm (NH)}  | \phi_n \rangle = 
E_n - i\sum_j\frac{\gamma_j}{2} \langle {L}_j^\dagger {L}_j \rangle_n$ are the complex, unperturbed eigenenergies of the NH Hamiltonian, possessing otherwise the same eigenbasis as $H_0$ thanks to the hypothesis Eq.~\eqref{e.commutation}.  

Comparing with standard non-degenerate perturbation theory \cite{Landau_BOOK_Quantum} it is immediate to verify that the first-order correction Eq.~\eqref{e.psi1} has the same form as that of the first-order perturbation to the eigenstates in closed systems when replacing $H_0 \to {H}_0^{\rm (NH)}$. The same applies as well to the first three lines of Eq.~\eqref{e.psi2}, but the fourth line is instead a new term purely introduced by dissipation.

\section{Systems of light emitters}
\label{sec:emitters}

\subsection{Individual vs. collective spontaneous emission}

In the rest of our work we shall specialize our attention to systems of light emitters -- such as real or artificial atoms undergoing spontaneous emission -- described as dissipative ensembles of two-level systems or qubits, with ground state $|g\rangle$ and excited state $|e\rangle$. They can be described in terms of spin operators ${\bm S}_i$ $(i = 1, ...., N)$, with corresponding spin states $|g\rangle = |\downarrow\rangle$ and $|e\rangle = \uparrow\rangle$ along the quantization axis $z$. A useful quantity in the rest of this work will be the collective spin operator ${\bm J} = \sum_i {\bm S}_i$. 
In the following we will assume the unperturbed Hamiltonian $H_0(\{ \bm S_i \})$ to be U(1) symmetric, namely conserving the collective spin component $J^z$, $[H_0, J^z]=0$, or, equivalently, the number of excited emitters. 

The coupling to the environment will be described in terms of spontaneous emission of photons, in the two limits of 1) individual spontaneous emission and 2) collective spontaneous emission:
\begin{enumerate}
\item \emph{individual emission}: in this case the jump operators are single-spin lowering operators $L_j \to S_i^-$, describing emitters which are all coupled to different light modes. If we assume the decay rates  to be uniform, $\gamma_j = \gamma$, we have that 
$\sum_j \gamma_j L_j^\dagger L_j$ reduces to   $\gamma \sum_i S_i^+ S_i^- = \gamma ( J^z+N/2 ) $, therefore commuting with the unperturbed Hamiltonian;  
\item \emph{collective emission}: in this case the jump operator is unique, $L_j \to J^- = \sum_i S_i^{-}$, describing emission of all emitters into the same light mode. In this case $\sum_j \gamma_j L_j^\dagger L_j = \frac{\Gamma}{N} J^+ J^- = \frac{\Gamma}{N} \left [ {\bm J}^2 - (J^z)^2 + J^z \right ] $. This operator commutes with the unperturbed Hamiltonian when the latter conserves not only the total magnetization, but also the collective spin length, namely if $[{\bm J}^2,H_0] = 0$. 
\end{enumerate}

\begin{figure*}[ht!]
  \centering
  \includegraphics[width=0.7\textwidth]{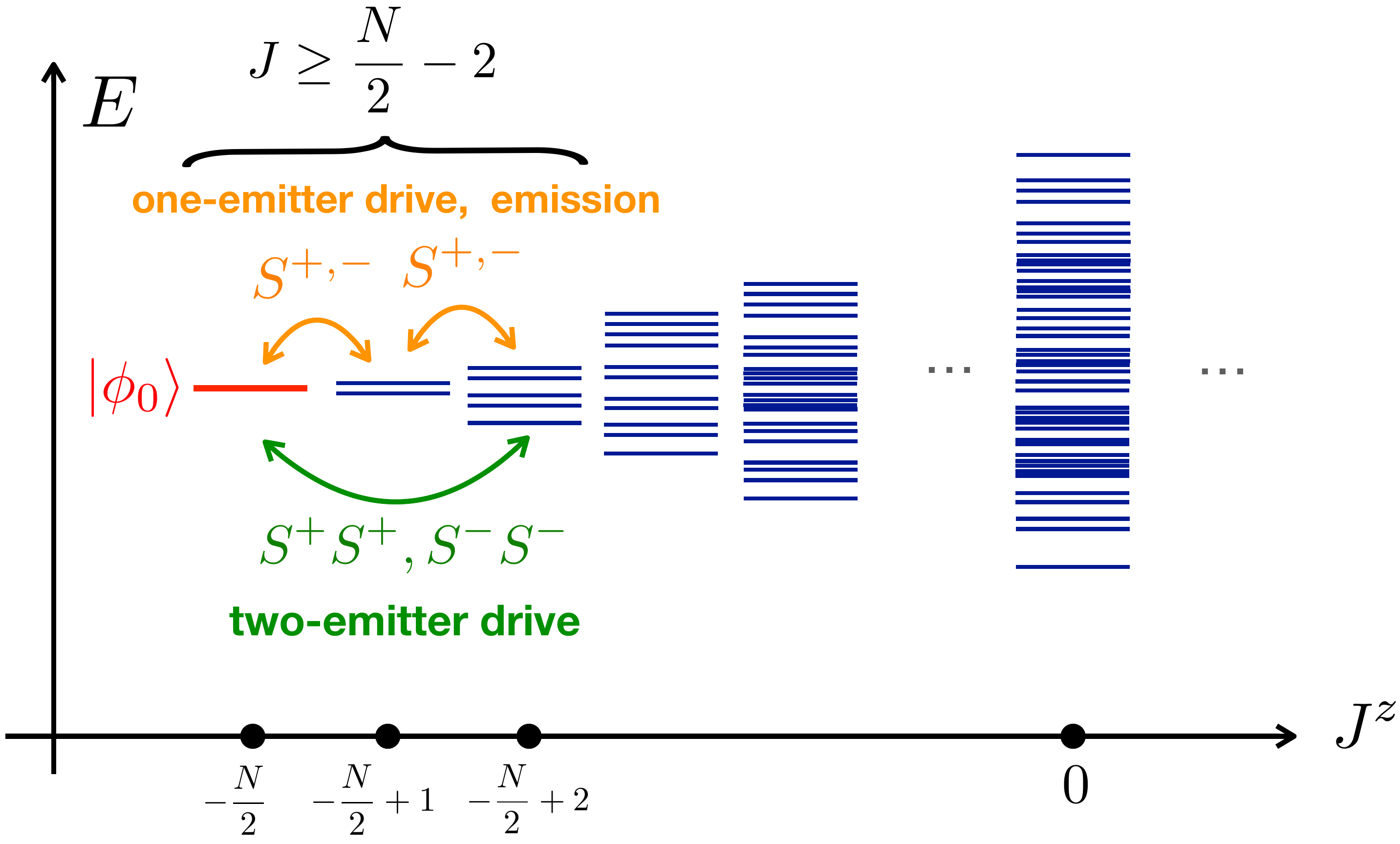}
  \caption{Sketch of perturbation theory from the unperturbed $|\phi_0\rangle = |\downarrow \downarrow ... \downarrow\rangle$ state. We picture here a generic spectrum of a $U(1)$-symmetric Hamiltonian, whose eigenstates can be labeled by the quantum number $J^z$. The two-emitter drive to first order, or the one-emitter drive to second order, connect the unperturbed state, with $J^z = -N/2$ and maximum spin length ${\bm J}^2 = J(J+1)$ and $J=N/2$, with states with $J^z = -N/2+2$ and spin length $J \geq N/2-2$. Hence the perturbed state is a superposition of highly atypical, large-spin Hamiltonian eigenstates.}
  \label{fig:spectrum}
\end{figure*}

For both of the above cases of interest to this work, the pure state $|\phi_0\rangle = |gg... g\rangle  = |\downarrow \downarrow ... \downarrow \rangle$ is the unique steady state of the unperturbed dissipative dynamics. Moreover both cases satisfy the assumption of the previous Sec.~\ref{s.special} (adding the extra condition $[{\bm J}^2,H_0] = 0$ in case 2), which leads to the explicit expressions for the state perturbation as in Eqs.~\eqref{e.psi1}-\eqref{e.psi2}. This situation lends itself to a rather interesting interpretation: the perturbed, pure steady state of the dissipative evolution can be viewed as the perturbed version of the Hamiltonian eigenstate $|\phi_0\rangle$ induced by the Hamiltonian $H_1$ to first order, and by the $H_1$ and the jump operators $L_j, L_j^\dagger$ to second order, namely as the perturbation of a (generically) excited Hamiltonian eigenstate. The perturbation operators $H_1$ and $L_j, L_j^\dagger$ will mainly admix $|\phi_0\rangle$ with other excited states in the same energy range, but with a fundamental restriction as we will highlight below. 

\subsection{Extremal properties of the unperturbed and perturbed state}
\label{s.extremal}

A relevant $H_1$ perturbation to $U(1)$ symmetric Hamiltonians breaks the U(1) symmetry, namely the conservation of the number of excited emitters. This implies that the perturbation corresponds generically to \emph{driving}, which we will consider as having two different forms (see sketch in Fig.~\ref{fig:sketch}): 1) \emph{two-emitter drive} exciting/de-exciting pairs of emitters together, of the kind $S_i^+ S_j^+ + S_i^- S_j^-$ (for $i\neq j$) where $S_i^{\pm} = S_i^x \pm i S_i^y$ are raising/lowering spin operators; and 2) \emph{single-emitter drive},  of the kind $S^+_i + S^-_i$. The two-emitter drive will be discussed in Secs.~\ref{sec:two_emitter}, while the single-emitter drive will be discussed in Sec.~\ref{sec:drivenDicke}. 

  At this point, it is crucial to understand the special role played by the unperturbed state $|\phi_0\rangle$ within perturbation theory. Indeed this state is extremal in that it maximizes (in absolute value) the projection $J^z$, and hence the collective spin length ${\bm J}^2= J(J+1)$ with $J=N/2$. All the operators associated to single-emitter and two-emitter drives, as well as the jump operators $L, L^\dagger$ related to emission, can only connect the $|\phi_0\rangle$ state to states with $J = N/2-2$ (to first order in the two-emitter drive operators, and to second order in the single-emitter drive and jump operators) -- see sketch in Fig.~\ref{fig:spectrum}. This implies that, to lowest order in perturbation theory, the perturbed state has only support on high-spin Hamiltonian eigenstates, representing a highly atypical portion of the spectrum, particularly so when the unperturbed Hamiltonian $H_0$ is not integrable. In the latter case, we can assume that the total spin length is not a good quantum number, and that the Hamiltonian spectrum in the energy range of the unperturbed steady state is composed mainly of typical states, generically displaying a small average length of the collective spin, namely $\langle {\bm J}^2 \rangle \sim O(N)$.
  
  This aspect suggests that, regardless of the details of the Hamiltonian, the nature of the perturbed steady state will be close to that obtained for models with long-range interactions (both in $H_0$ as well as in $H_1$), whose eigenstates possess a well defined total spin length  ${\bm J}^2$ with small or no fluctuations. In the rest of this work we will work with collective-spin Hamiltonians, whose choice is not only convenient technically (see below), but also justified by the above remarks.

\section{Collective-spin properties and spin squeezing}
\label{sec:collective_spin}

Before we apply pure-state perturbation theory to specific examples, it is important to introduce the quantities that we shall use to characterize the quantum correlations induced by the perturbation in the steady state of the system. 

The unperturbed steady state of our interest $|\phi_0\rangle = |\downarrow \downarrow ... \downarrow \rangle$  is an example of a \emph{coherent spin state} (CSS). The collective spin of this state points along the $z$ axis, $\langle J^z \rangle = - \frac{N}{2}$, and the transverse spin components have equal uncertainties
\begin{equation}
 {\rm Var}(J^x) = {\rm Var}(J^y)=N/4~.
\end{equation}
In fact we can extend the above property by considering a generalized definition of the collective spin as having transverse, perpendicular components $J_1^\perp$ and $J_2^\perp$, such that 
$[J_1^\perp,J_2^\perp] = iJ^z$,  in the form 
\begin{align}
J_1^{\perp} =& \sum_i \left ( \cos\theta_i S_i^x + \sin\theta_i S_i^y \right) \nonumber \\
=& \frac{1}{2} \sum_i \left(  S_i^+ e^{-i\theta_i} + S_i^- e^{i\theta_i}\right) ~,
\label{e.J1perp}
\end{align}
and
\begin{align}
J_2^{\perp} =& \sum_i \left ( -\sin\theta_i S_i^x + \cos\theta_i S_i^y \right) \nonumber \\
=& - \frac{1}{2} \sum_i \left(  S_i^+ e^{i\theta_i} + S_i^- e^{-i\theta_i}\right) ~.
\label{e.J2perp}
\end{align}
where $\theta_i$ are local angles. For a CSS we have that ${\rm Var}(J_1^{\perp}) = {\rm Var}(J_2^{\perp}) = N/4$, and in particular ${\rm Var}(J_1^{\perp}) {\rm Var}(J_2^{\perp}) = |\langle J^z \rangle|^2/4$, saturating the Heisenberg uncertainty relation for spin operators. As a consequence the CSS represents an example of a \emph{minimal uncertainty} state.  
 
As we shall see, Hamiltonian perturbations breaking its U(1) symmetry have the immediate effect of redistributing the uncertainties on the transverse components in the steady state, inducing \emph{squeezing} of the collective spin. 
Squeezing of a transverse collective spin component can be quantified by the squeezing parameter \cite{WinelandPRA1994}  
 \begin{equation} \label{eq:squeezing}
\xi^2_R = \frac{N\min_{\perp} {\rm Var}(J^\perp)}{\langle J^z \rangle^2}~.
\end{equation}
where the minimization $\min_{\perp}$ is made over all possible choices of the local angles $\theta_i$. If $\xi_R^2 < 1$ the collective spin exhibits squeezing, which is a fundamental form of quantum correlation. Indeed squeezed states are entangled \cite{Sorensen2001}; and their entanglement has an immediate metrological significance, making them more sensitive to rotations more than any coherent spin state \cite{WinelandPRA1994}. 

Because the collective spin components must satisfy the Heisenberg uncertainty relation  ${\rm Var}(J_1^{\perp}) {\rm Var}(J_2^{\perp}) \geq |\langle J^z \rangle|^2/4$, if one component (e.g. $J_1^\perp$) acquires the minimum variance $\min_{\perp} {\rm Var}(J^\perp)$ and is squeezed (namely $\xi_R^2<1$), the other one (e.g. $J_2^\perp$) must be \emph{anti-}squeezed (namely $\xi_R^2>1$ for it), and in particular it has the property of exhibiting the largest variance among all the collective-spin components transverse to the average orientation. The anti-squeezed component captures therefore the strongest form of spin-spin correlations appearing in the system. To quantify the quantum nature of these correlations, one can introduce the quantum Fisher information (QFI) associated with $J_2^\perp$, which is most generally defined for a state $\rho = \sum_k p_k |k\rangle \langle k|$ (with $|k\rangle$ forming an orthonormal basis) as \cite{Braunstein94}
\begin{equation}
{\rm QFI}(J_2^\perp) = 2 \sum_{kl} \frac{(p_k - p_l)^2}{p_k-p_l} | \langle k | J_2^\perp | l \rangle |^2~.
\end{equation}
In particular the QFI bounds the uncertainty with which one can estimate the angle $\phi$ of a rotation $e^{-i\phi J^\perp_2}$, generated by the operator $J_2^\perp$, by making arbitrary measurements on the state, $(\delta\phi)^2 \geq 1/{\rm QFI}(J_2^\perp)$. 
The {\rm QFI} satisfies the inequality (descending from the above bound and the fact that ${\rm QFI}(J_2^\perp) \leq 4 {\rm Var}(J_2^{\perp})$ \cite{PezzeRMP2018}):
\begin{equation}
4 {\rm Var}(J_1^{\perp}) {\rm Var}(J_2^\perp) \geq {\rm Var}(J_1^{\perp}) {\rm QFI}(J_2^\perp) \geq |\langle J^z \rangle|^2
\label{e.ineq_chain}
\end{equation}
implying that 
\begin{equation}
\xi_R^{-2} \leq   {\rm QFI}(J_2^\perp)/N~.
\end{equation}
If the squeezed state is a state of minimal uncertainty, this means that the inequality chain Eq.~\eqref{e.ineq_chain} collapses to an equality, and in particular that $\xi_R^{-2}  =   {\rm QFI}(J_2^\perp)/N$. For such a state squeezing is the optimal metrological resource, namely rotations can be best estimated by simply measuring the average orientation of the collective-spin operator.

\section{Two-emitter driving and individual emission: squeezing theorem}
\label{sec:two_emitter}

\subsection{Two-emitter driving}

The first example of perturbation that we shall consider is a most general, parity-conserving bilinear Hamiltonian perturbation $H_1$ in the form
$H_1 = H_{1,d} +  H_{1,o}$ where  
\begin{equation}
H_{1,d} =  \sum_{lm} \left ( J_{lm}  S_l^+ S_m^- + {\rm h.c.} \right ) + H_{\rm diag}
\label{eq:H1d}
\end{equation}
is a $J^z$-conserving term, and 
\begin{equation}
H_{1,o} = \sum_{lm} \left ( K_{lm} S_l^+ S_m^+  +  {\rm h.c.} \right )~
 \label{eq:H1o}
\end{equation} 
is a $J^z$-non-conserving one.
Here $H_{\rm diag}$ is an arbitrary Hamiltonian which is diagonal on the eigenbasis of the $S_i^z$ operators. The perturbation $H_{1,o}$ contains two-emitter driving terms, which are essential to perturb the steady state. 

Indeed, given that $[H_0,J^z]=0$, the Hamiltonian eigenstates $|\phi_n\rangle$ can be chosen to be also eigenstates of $J^z$. In particular $|\phi_0\rangle$ is the only state in the $J^z = -N/2$ sector. Given that $ H_{1,d}$ cannot couple different $J^z$ sectors, we have that 
\begin{equation}
\langle \phi_n | H_{1,d} | \phi_0 \rangle \sim \delta_{n0}~.
\end{equation}
Hence the state $|\phi_0\rangle$ can only be perturbed by the $H_{1,o}$ operator. Within that operator the terms of the kind $S_l^+ S_m^+ $ are the only ones that do not annihilate the $|\phi_0\rangle$ state, and they map that state onto states belonging to the $J^z = -N/2+2$ sector, that we will denote as ${\cal S}_{-N/2+2}$ in the following.

Since the perturbation $H_1$ introduces emitter-emitter couplings, it can by itself give rise to quantum correlations in the steady state. Hence we will consider the case of individual (i.e. uncorrelated) spontaneous emission $L_j \to S_i^-$ with a uniform rate $\gamma$ for all emitters. This coupling to the environment has clearly the tendency to decorrelate the emitters. The form of quantum correlations stabilized in the steady state will therefore be of a kind which is robust to a decorrelating environment.

\subsection{Squeezing theorem}

Applying the formulas of Sec.~\ref{s.special}, the first-order perturbation to the steady state takes the form 
\begin{equation}
|\psi_1\rangle = \sum_{n \in {\cal S}_{-N/2+2}} \frac{\langle \phi_n | \sum_{lm} K_{lm} S_l^+ S_m^+ | \phi_0 \rangle}{E_0 - E_n + i\gamma } |\phi_n\rangle,
\label{e.ps1}
\end{equation}
since
\begin{equation}
\frac{1}{2} \langle \phi_n | \sum_i S_i^+ S_i^- |\phi_n \rangle = \frac{1}{2} (-N/2 + 2 + N/2) = 1~.
\end{equation}
As a consequence
\begin{align}
 \langle (J^{\perp})^2  \rangle' = 
 \frac{N}{4} + \lambda \left ( \langle \phi_0 | (J^{\perp})^2 | \psi_1 \rangle + {\rm c.c.} \right ) + O(\lambda^2), 
\end{align}
where we introduced the short-hand notation for the expectation values on the perturbed state $\langle (...) \rangle' = \langle \phi_0' | (...) |\phi_0' \rangle$. The matrix element $\langle \phi_0 | (J^{\perp})^2 | \psi_1 \rangle$ can be written as
\begin{widetext}
\begin{equation}\label{eq:Fequation}
\langle \phi_0 | (J^{\perp})^2 | \psi_1 \rangle = - \frac{1}{4} \sum_{n \in {\cal S}_{-N/2+2}}  \frac{\sum_{lm} \sum_{ij} e^{i(\theta_i+\theta_j)} \langle \phi_n |  K_{lm} S_l^+ S_m^+ | \phi_0 \rangle  \langle \phi_0 | S_i^- S_j^- |\phi_n \rangle }{E_n- E_0 - i \gamma
} = -F(\{\theta_i\}) ~.
\end{equation}
\end{widetext}
Given that 
\begin{equation}
\langle  J^z  \rangle' = - \frac{N}{2} + O(\lambda^2),
\end{equation}
the squeezing parameter to first order takes the value 
\begin{equation}\label{eq:squeezing_equation}
\xi_R^2 = 1  - \frac{8\lambda}{N} ~ {\rm Re}[F(\{\theta_i\})] + O (\lambda^2)~.
\end{equation}
The above results implies the following

\begin{theorem}\label{theorem1}
The steady state of an ensemble of emitters with U(1)-symmetric unperturbed Hamiltonian subject to individual emission -- whose unique steady state is 
$|\phi_0\rangle = |\downarrow \downarrow ... \rangle$ -- is perturbed to a spin-squeezed state (within first-order perturbation theory) by a perturbation of the form of Eq.~\eqref{eq:H1o} iff
\begin{equation}
\max_{\{\theta_i\}} {\rm Re}[ F(\{\theta_i\})] > 0~
\label{eq:condition_maxsqueezing}
\end{equation}
where the function $F$ is defined in Eq.~\eqref{eq:Fequation}. 
\end{theorem}

The angles $\{\theta_i\}$ maximizing the function ${\rm Re}[F]$ define the generalized collective-spin component $J^{\perp}_1$ which is maximally squeezed; and, in turn, the angles  $\{\theta_i + \pi/2\}$ define the anti-squeezed component $J_2^{\perp}$, namely the local spin components $- \sin(\theta_i) S_i^x + \cos(\theta_i) S_i^y$ which are most strongly correlated with one another between different sites. 

If the condition Eq.~\eqref{eq:condition_maxsqueezing}, defining the squeezed transverse component $J^{\perp}_1$,  is satisfied, then the perpendicular component $J^{\perp}_2$ is automatically anti-squeezed since $\langle \phi_0 | (J_1^\perp)^2 | \psi_1\rangle = - \langle \phi_0 | (J_2^\perp)^2 | \psi_1\rangle$, and therefore 
\begin{equation}
\langle (J^\perp_2)^2  \rangle' = \frac{N}{4} + 2 \lambda \max_{\{ \theta_i\}} {\rm Re}[F({\theta_i}]+ O(\lambda^2).
\end{equation}
As a consequence, the perturbed state has the property
 \begin{equation}
\langle  (J^\perp_1)^2  \rangle'  \langle (J^\perp_2)^2  \rangle' = \frac{N^2}{16} + O(\lambda^2) = \frac{|\langle J^z \rangle'|^2}{4} + O(\lambda^2) 
\end{equation}
namely it is a \emph{minimal uncertainty} state, for which squeezing is the optimal metrological property. 

The above theorem extends to dissipative steady states a simple theorem recently put forward by one of the authors for Hamiltonian ground states \cite{Roscildeetal2022}. The present result highlights the  importance of spin squeezing as the first form of quantum correlation that can be stabilized when perturbing the factorized stead state $|\phi_0\rangle$ with a bilinear perturbation (a two-emitter drive). 
The assumption of having a bilinear perturbation in the spin operators is not a fundamental limitation. 
If we had included in $H_1$ an off-diagonal perturbation which is linear in the spin operators, \emph{e.g.} proportional to $J^x$ or $J^y$,  this would not change the above results within first-order perturbation theory. Indeed, due to the structure of the squeezing parameter Eq.~\eqref{eq:squeezing}, linear operators in the spins cannot give contributions to first order, because they cannot compensate the effect of the bilinear operator $(J^{\perp})^2$ on the unperturbed state (see Eq.~\eqref{e.psi1}). On the other hand they can perturb the state to second order, as we will discuss in Sec.~\ref{sec:drivenDicke}.

We would like to stress that the prediction of quantum correlations appearing at first order in perturbation theory is a reliable result of pure-state perturbation theory, since, as we discussed in Sec.~\ref{sec:pure}, the purity of the state is only altered to second order in the perturbation.

In the following section we will test the predictions of pure-state perturbation theory and the relevance of spin squeezing in two examples widely studied in the recent literature on dissipative quantum spin models.

\section{Application to collective spin Hamiltonians}
\label{sec:applications}

In the following we will consider collective-spin Hamiltonians of the form
\begin{equation}
    {H} = \sum_{\alpha=x,y,z} \left [  \frac{\mathcal{J}_\alpha}{N} (J^\alpha)^2 + h_\alpha J^\alpha\right ],
\end{equation}
which can  be  decomposed into an unperturbed, U(1)-symmetric part, and a symmetry-breaking perturbation. 
Specifically, we will discuss the collective XYZ model \cite{HuybrechtsPRB20} and the collective transverse-field Ising model \cite{OverbeckPRA17} with dissipation in the form of individual emission with uniform decay rates. These models are particularly suitable for a test of our perturbative results, since they admit an efficient numerical solution thanks to the permutational symmetry of their Hamiltonians \cite{ShammahPRA98}, offering a valuable benchmark. For these numerical simulations we resort to the implementation of the permutational invariant solver in QuTiP \cite{qutip1, qutip2}.

\subsection{Example I: the dissipative XYZ model}\label{sec:xyz}
We consider the collective (or infinite-dimensional) XYZ Heisenberg model \cite{HuybrechtsPRB20} with individual spontaneous emission, whose Hamiltonian can be written as
\begin{equation}
    H =  \frac{1}{N} \sum_{i,j}\left({\cal J}_x S_i^xS_j^x + {\cal J}_y S_i^yS_j^y + {\cal J}_z S_i^zS_j^z \right),
    \label{e.HXYZ}
\end{equation}

where the sums $\sum_{i(j)}$ run over the $N$ emitters. 
This Hamiltonian -- both in its collective-spin form, as well as when cast on finite-dimensional lattices -- has been the subject of many theoretical studies in the recent past \cite{LeePRL13,JinPRX16,RotaPRB17,RotaNJP18, HuybrechtsPRA19, HuybrechtsPRB20, Verstraelenetal2023}. 

Choosing $\mathcal{J}_x = \mathcal{J}_y$, one recovers a U(1)-symmetric XXZ model, conserving the magnetization $J^z$. The dissipative XXZ model clearly admits the state  
$|\phi_0\rangle = |\downarrow \downarrow ... \downarrow\rangle $ as its unique steady state. It is useful to recast the latter state as a Dicke state $|\phi_0\rangle=|J=N/2, M=-N/2\rangle$ with maximal spin length ${\bm J}^2 = J(J+1)=N/2(N/2+1)$ and $J^z =M$. 

The Hamiltonian Eq.~\eqref{e.HXYZ} can be rewritten in the form
\begin{equation}
    {H} = H_0 +\lambda H_1,
\end{equation}
with 
\begin{equation}
    H_0 = \frac{\mathcal{J}}{N}\left[(J^x)^2 + (J^y)^2\right] + \frac{\mathcal{J}_z}{N} (J^z)^2,
\end{equation}
and
\begin{align}
    \lambda H_1  = \frac{ \delta {\cal J}}{2N} \left [ (J^+)^2 + (J^-)^2 \right ]  
    \end{align}
where we have introduced the symbols
\begin{equation}
{\cal J} = \frac{{\cal J}_x+{\cal J}_y}{2}~~~~~~ {\delta \cal J} = \frac{{\cal J}_x-{\cal J}_y}{2}~.
\end{equation}

Given the permutation invariance of the Hamiltonian (including the perturbation), we can choose uniform angles $\theta_i = \theta$ in the definition of the transverse spin components Eqs.~\eqref{e.J1perp} and \eqref{e.J2perp} without loss of generality. In particular the permutationally invariant perturbation will couple the Dicke state $|\phi_0\rangle = |J=N/2, M=-N/2\rangle$ to another Dicke state only, namely $|\phi_2\rangle = |J=N/2, M=-N/2+2\rangle$ via a double spin flip. This means that the sum of Eq.~\eqref{eq:Fequation} collapses to only one state. 
Moreover Dicke states are all eigenstates of the all-to-all XXZ Hamiltonian with eigenvalues 
\begin{align}
E(J,M) =& \langle J, M | H_0 | J, M \rangle, \\
= & \frac{\cal J}{N}  J(J+1) +  \frac{1}{N} ({\cal J}_z - {\cal J} )M^2  ,
\end{align}
so that the only relevant transition energy is $E_2 - E_0 =  2\left({\cal J} - {\cal J}_z \right)\left(N - 2\right)/N$.

Moreover we get 
\begin{align}
\langle \phi_2 | H_1 | \phi_0 \rangle & =   \frac{\delta \cal J}{2N} \left\langle \phi_2 \right\vert(J^+)^2\left\vert\phi_0  \right\rangle, \\
& =  \frac{\delta \cal J}{2} \sqrt{2\left(N-1\right)/N}. \nonumber
\end{align}
As a consequence the function $F\left(\{\theta\}\right)$ from theorem 1 takes the form
\begin{align}\label{eq:complexreprF}
F(\{\theta\}) =& \frac{1}{8}\frac{\delta {\cal J}(N-1)e^{i2\theta}}{\left({\cal J} - {\cal J}_z \right)\left(N - 2\right)/N - i\gamma},\\
=& \frac{\alpha e^{i2\theta}}{\beta - i\gamma}. 
\end{align}
where we introduced the symbols 
\begin{align}
\alpha & = \frac{N-1}{4} ~ \delta {\cal J} \\
\beta  & = \frac{2(N-2)}{N} \left({\cal J} - {\cal J}_z \right)~.
\end{align}
In terms of these symbols the real part of $F$ reads 
\begin{equation}\label{eq:ReF}
{\rm Re}[F(\{\theta\})] = \frac{\alpha\left(\beta\cos2\theta - \gamma\sin2\theta\right)}{\beta^2 + \gamma^2}.
\end{equation}

\begin{figure}[ht!]
  \centering
  \includegraphics[width=0.5\textwidth]{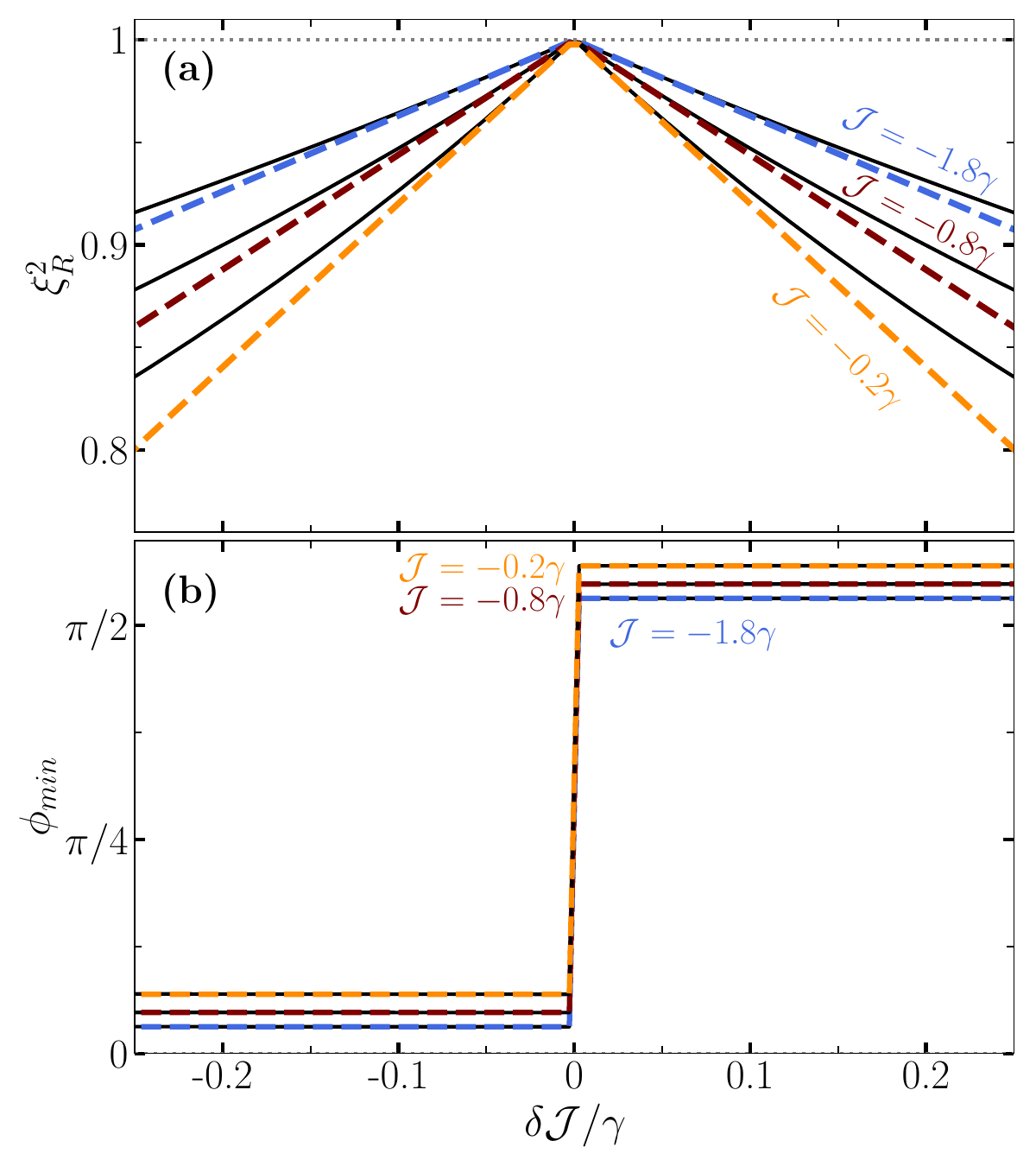}
  \caption{(a) Minimal spin squeezing parameter $\xi_{\rm R}^2$ as a function of $\delta \mathcal{J}$ for the exact solution (full black lines) and the perturbative solution (dashed colored lines). This corresponds to a perturbation perpendicular to the $\mathcal{J}$ axis (see also Fig.~\ref{fig:PhDsketch_XYZ}). Parameters used are $N=20$, $\mathcal{J}_z  = \gamma$ , and $\mathcal{J} = -1.8\gamma, -0.8\gamma, -0.2\gamma$ (resp. blue, dark red, and orange, as indicated on the figure).   
  (b) The angle $\phi_{min}$ for which the spin squeezing parameter is minimal (see panel (a)). Same parameters as panel (a) (from top to bottom; $\mathcal{J} = -0.2\gamma, -0.8\gamma, -1.8\gamma$).
  }
  \label{fig:xyz}
\end{figure}

\subsubsection{Onset of squeezing in the perturbed state}

The presence of squeezing in the steady state is then governed by condition \eqref{eq:condition_maxsqueezing}, and depends on the angle $\theta$. The angle $\theta$ extremizing  ${\rm Re}[F(\{\theta_i\})]$ is given by (see appendix \ref{app:ReFex})
\begin{equation}\label{eq:theta_ex}
\theta_{ex.} = \frac{1}{2}\tan^{-1}\left(-\frac{\gamma}{\beta}\right) + k\frac{\pi}{2} \quad k\in\mathbb{Z}
\end{equation}
and it exhibits a subtle dependence on both the Hamiltonian parameters and the dissipation strength that we will discuss in Sec.~\ref{s.angleXYZ}. 

For the extremal angle $\theta_{\rm ex}$ the real part of the $F$ function takes the form  (see appendix \ref{app:ReFex})
\begin{equation}\label{eq:ReFextremalvalue}
    {\rm Re}\left[F(\left\{\theta_{\rm ex}\right\})\right] 
    = \frac{\alpha\cos(k\pi)}{\sqrt{\beta^2 + \gamma^2}}.
\end{equation}
The sign of this function (determining the existence of squeezing) depends on the integer $k$ and on the sign of $\alpha$, but it does not depend on the sign of $\beta$. Yet, whenever the perturbation breaks the U(1) symmetry, namely $\alpha \neq 0$, one can always find an extremizing angle for which ${\rm Re}(F)>0$: this is achieved by choosing $k$ even (odd) for $\alpha>0$  ($\alpha < 0$).  This implies that \emph{the U(1)-symmetry-breaking perturbation always induces squeezing in the steady state}, as per the above theorem. 
The above result is clearly exhibited in Fig. \ref{fig:xyz}, where we compare the squeezing parameter predicted by perturbation theory with the exact result for $N=20$ spins as an example. There we see that, as predicted by pure-state perturbation theory, the squeezing parameter $\xi_R^2$ decreases linearly from its unperturbed unit value. The deviation from the perturbative predictions can clearly be attributed to beyond-linear effects, as well as to the fact that the steady-state develops some finite entropy (albeit subextensive \cite{Huybrechts_in_preparation}), not accounted for by pure-state perturbation theory. 

The prediction of perturbation theory for the squeezing parameter agrees with the exact result over a larger interval of perturbation values when  $\mathcal{J} \ll \mathcal{J}_z$ and $\mathcal{J} \gg \mathcal{J}_z$, namely far away from the SU(2)-symmetric point (for the Hamiltonian) ${\cal J}_x = {\cal J}_y = {\cal J}_z$. To understand this result it is important to remind that, for a critical value of the perturbation $\delta {\cal J}$ (dependent on the value of ${\cal J}/\gamma$), the steady-state phase diagram features a phase transition \cite{HuybrechtsPRB20} from a paramagnetic phase -- continuously connected with the unperturbed U(1)-symmetric steady state; to a ferromagnetic phase, which develops long-range correlations for one spin component in the $xy$ plane (see the sketch in Fig.~\ref{fig:PhDsketch_XYZ}, and Sec.~\ref{s.angleXYZ} for further discussion). The critical value of $\delta{\cal J}$ is minimal (in absolute value) for ${\cal J}={\cal J}_z$ and it grows upon increasing the difference between ${\cal J}$ and ${\cal J}_z$. The predictions of perturbation theory break down upon approaching the phase transition, and this occurs for larger  $\delta {\cal J}$ values the larger the anisotropy between ${\cal J}$ and ${\cal J}_z$. 
Nonetheless, even when perturbing the steady state away from the SU(2) symmetric point, perturbation theory correctly captures the onset of squeezing, showing that this is a universal feature of the paramagnetic phase close to the U(1)-symmetric line ${\cal J}_x = {\cal J}_y$. As we will discuss in the next section, pure-state perturbation theory also predicts correctly which collective-spin component is squeezed and which one is anti-squeezed; and in this sense, it provides insight into important features of the steady-state phase diagram, even going beyond its strict range of applicability.

\begin{figure*}[ht!]
  \centering
  \includegraphics[width=0.8\textwidth]{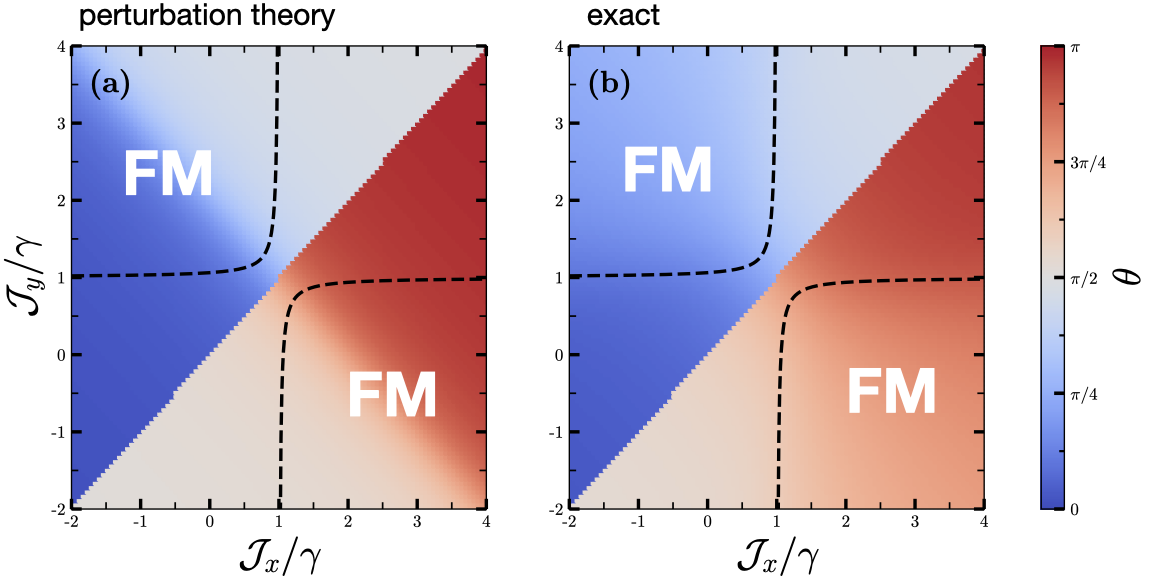}
  \caption{Optimal angle $\theta$ in the $(x,y)$-plain that minimises the spin squeezing parameter in the XYZ model as predicted by (a) the perturbative approach and (b) the exact solution for a system with $N = 20$ spins and $\mathcal{J}_z = \gamma$. Black dashed lines show the mean-field phase transition between the paramagnetic phase and ferromagnetic phase -- corresponding to the exact solution for the all-to-all connected model in the thermodynamic limit \cite{HuybrechtsPRB20, CarolloPRL21}.
  }
  \label{fig:XYZphase}
\end{figure*}

\subsubsection{Angle of squeezing vs. anti-squeezing; insight into the ferromagnetic phases}
\label{s.angleXYZ}

Eq.~\eqref{eq:theta_ex} from perturbation theory gives an explicit prediction of the extremal angle for squeezing and anti-squeezing as a function of the Hamiltonian parameter and the dissipation strength. The extremal squeezing angle $\theta_{\rm ex}$ is plotted in Fig.~\ref{fig:XYZphase} across the steady-state phase diagram, as a function of the ratios ${\cal J}_x/\gamma$ and ${\cal J}_y/\gamma$ at fixed ${\cal J}_z/\gamma=1$, comparing pure-state perturbation theory with the exact result from Ref.~\cite{HuybrechtsPRB20}. We observe that the subtle dependence of the angle on both the Hamiltonian parameters and the dissipation strength is correctly captured  by perturbation theory within its strict regime of applicability, namely within the paramagnetic phase continuously connected with the unperturbed limit ${\cal J}_x = {\cal J}_y$. 

\begin{figure}[ht!]
\begin{center}
\includegraphics[width=0.9\columnwidth]{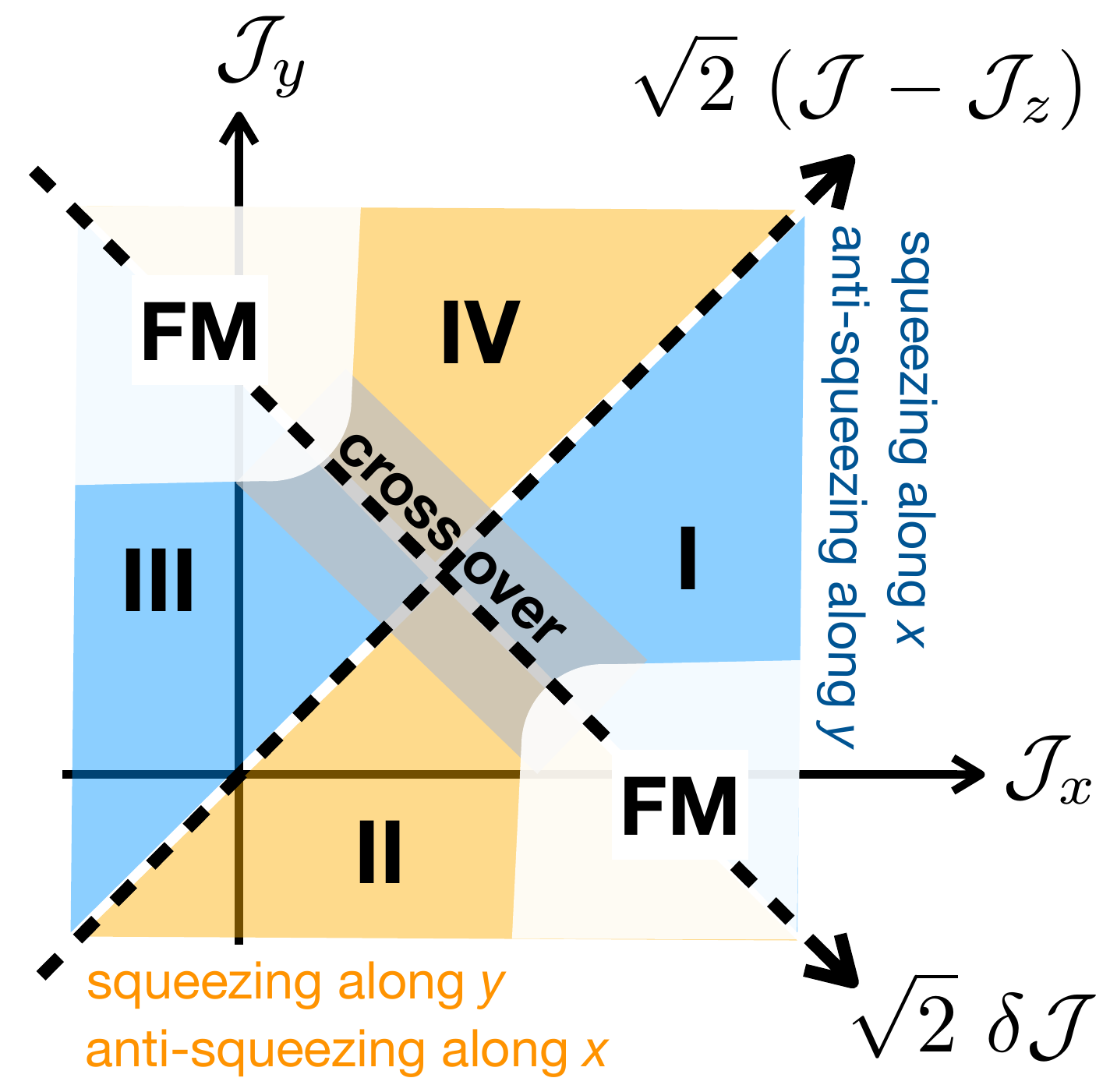}
\caption{Sketch of the steady-state phase diagram of the dissipative XYZ model, indicating the squeezed/anti-squeezed collective-spin components predicted by perturbation theory.}
\label{fig:PhDsketch_XYZ}
\end{center}
\end{figure}

In order to make sense of the angle map of Fig.~\ref{fig:XYZphase}, it is instructive to take the limite of a negligible dissipation, i.e. to consider $|\mathcal{J -J}_z| \gg \gamma$. In this limit the ratio $\gamma/\beta$ becomes negligible, and therefore one obtains ${\rm Re}[F(\theta)] \approx \alpha \cos(2\theta)/\beta$, which for $\alpha/\beta>0$ is maximized by $\theta=0 ~({\rm mod}~\pi)$ and for $\alpha/\beta<0$ it is maximized by $\theta = \pi/2 ~({\rm mod}~\pi)$. 
The two conditions, $\alpha/\beta>0$ and $\alpha/\beta<0$, divide the $({\cal J}_x,{\cal J}_y)$ plane into four quadrants, which are oriented with the $\pi/4$-rotated axes 
$\delta {\cal J}$ and ${\cal J}-{\cal J}_z$.  
The condition $\alpha/\beta>0$  is met when $\delta {\cal J} >0$ and ${\cal J}> {\cal J}_z$, defining the quadrant I;  or $\delta {\cal J} <0$ and ${\cal J}< {\cal J}_z$, defining the quadrant III. In these two quadrants the maximum squeezing angle is therefore $\theta=0 ~({\rm mod}~\pi)$. The angle $\theta=\pi/2 ~({\rm mod}~\pi)$ is instead the maximum squeezing angle in the quadrant II: $\delta {\cal J} >0$ and ${\cal J}< {\cal J}_z$; and  quadrant IV: $\delta {\cal J} < 0$ and ${\cal J} > {\cal J}_z$. The four quadrants with $\theta = 0 ~({\rm mod}~\pi)$ and $\theta = \pi/2 ~({\rm mod}~\pi)$ are clearly visible in Fig.~\ref{fig:XYZphase}(a); and they are repeated in Fig.~\ref{fig:PhDsketch_XYZ} for clarity. 
Introducing a finite dissipation $\gamma$ has then the main effect of rounding off the transition between the two values of $\theta =0 ~({\rm mod}~\pi)$ or $\pi/2 ~({\rm mod}~\pi)$ when moving above or below the ${\cal J}_x= {\cal J}_y$ line.  

The angle $\theta$ shown in the figure determines the maximally squeezed component of the collective spin, but correspondingly the angle $\theta +\pi/2 ~({\rm mod}~\pi)$ determines the anti-squeezed spin component, developing the strongest ferromagnetic correlations. These correlations become long-ranged across the paramagnetic-to-ferromagnetic transition, whose description is clearly beyond the scopes of perturbation theory. Yet the nature of the ferromagnetic correlations can be understood to a large extent from perturbation theory. 

First of all, it is remarkable to see that the dominant correlations are in the $xy$ plane, and they are ferromagnetic \emph{regardless} of the signs of the Hamiltonian couplings ${\cal J}_x$ and ${\cal J}_y$. Moreover the spin component exhibiting the strongest correlations does not correspond necessarily to the spin component with the largest Hamiltonian coupling. Perturbation theory predicts that the strongest ferromagnetic correlations are exhibited by the $x$ spin components in quadrants I and III, and by the $y$ spin component in quadrants II and IV; while the strongest coupling in the Hamiltonian is ${\cal J}_x$ in quadrants II and III, and ${\cal J}_y$ in quadrants I and IV. Clearly the onset of ferromagnetism in the steady state cannot be easily understood in terms of energy extremization.

On the other hand, the pure-state perturbation picture predicts that the perturbed state has the simple form of a superposition of two Dicke states:
\begin{align}
|\phi_0'\rangle \approx & |J=N/2,M=-N/2 \rangle \nonumber \\
&+ \zeta | J=N/2,M=-N/2+2 \rangle  
\end{align}
such that 
\begin{align}
\langle \phi_0' | S_i^x S_j^x | \phi_0' \rangle & = \zeta'  \nonumber \\
\langle \phi_0' | S_i^y S_j^y | \phi_0' \rangle & = -\zeta' 
\end{align}
where $\zeta' =  \zeta / \sqrt{2N(N-1)}$~. Given that 
\begin{equation}
\zeta = -\frac{\sqrt{ 2N(N-1)}}{4(N-2)} \frac{\delta {\cal J}}{{\cal J} - {\cal J}^z}
\end{equation}
we obtain positive, ferromagnetic correlations (and anti-squeezing) for the $x$ spin component when ${\delta {\cal J}} < 0, {{\cal J}-{\cal J}^z} > 0$ or ${\delta {\cal J}} >0, {{\cal J}-{\cal J}^z}<0$
(quadrants II and IV respectively), while negative correlations (ans squeezing) appear for the $y$ spin component. On the other hand, for ${\delta {\cal J}} >0, {{\cal J}-{\cal J}^z}>0$ and  ${\delta {\cal J}} <0, {{\cal J}-{\cal J}^z}<0$ (quadrants I and III respectively) ferromagnetic correlations appear for the $y$ spin component etc. This is in agreement with the previous analysis of the extremizing angle.  The picture of the correlation structure is summarized in Fig.~\ref{fig:PhDsketch_XYZ}.

\begin{figure}[t]
  \centering
  \includegraphics[width=0.5\textwidth]{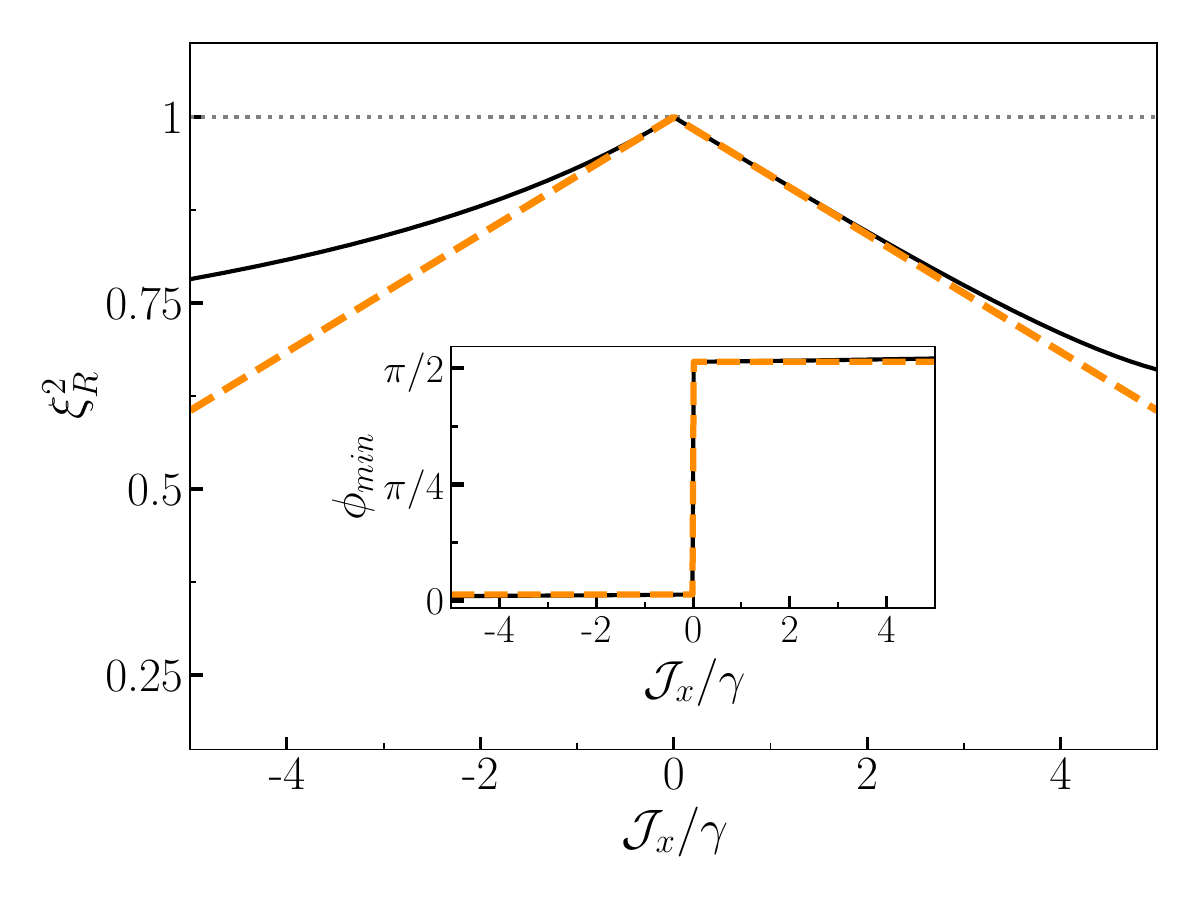}
  \caption{Minimal spin squeezing parameter $\xi_{\rm R}^2$ in the dissipative TFI model as a function of $\mathcal{J}_x$ for the exact solution (full black line) and the perturbative solution (dashed orange line).  Inset: the angle $\phi_{min}$ for which the spin squeezing parameter is minimal. Parameters used are $N=20$,$\Delta = -6\gamma$ (as indicated on the figure). 
  }
  \label{fig:tfi}
\end{figure}

\subsection{Example II: the dissipative transverse field Ising model}\label{sec:tfi}

In the following we discuss a second model with two-emitter driving, namely the transverse-field Ising (TFI) model with collective (or infinite-dimensional) interactions and spontaneous local emission along the field axis \cite{OverbeckPRA17}, whose Hamiltonian is given by
\begin{equation}
    H = \frac{{\cal J}_x}{N} \sum_{i,j}S_i^xS_j^x + \Delta \sum_i S_i^z~.
\end{equation}
The dissipation is the same as for the previous model, namely we have single-emitter jump operators $L_j \to  S_i^-$ with uniform single-emitter decay rate $\gamma$.

The Hamiltonian $H$ can be decomposed as a U(1)-symmetric unperturbed term and a symmetry-breaking perturbation ${H} = H_0 +\lambda H_1$, with
\begin{equation}
    H_0 = \Delta  J^z,
\end{equation}
and
\begin{align}
    \lambda H_1 & = \frac{{\cal J}_x}{N} \sum_{i,j}S_i^xS_j^x  \\
    & = \frac{{\cal J}_x}{4N} \left [  ( J^+)^2 + (J^-)^2 + J^+ J^- + J^- J^+ \right] \nonumber~.
\end{align}
The unique steady state $|\phi_0\rangle$ in the unperturbed limit $\lambda = 0$ is, as before, the coherent spin state aligned with the $-z$ axis. And, as in the previous example, it is connected by the perturbation $H_1$ to only one eigenstate of the unperturbed Hamiltonian, namely the Dicke state $|\phi_2\rangle = |J=N/2, M=-N/2+2\rangle$. Since Dicke states have the unperturbed energies $E(J=N/2,M) = \Delta M$,  the energy difference entering in the denominator of the first-order correction to the steady state, Eq.~\eqref{e.psi1}, is simply $E_2-E_0 = 2\Delta$.

\begin{figure*}[t]
  \centering
  \includegraphics[width=0.8\textwidth]{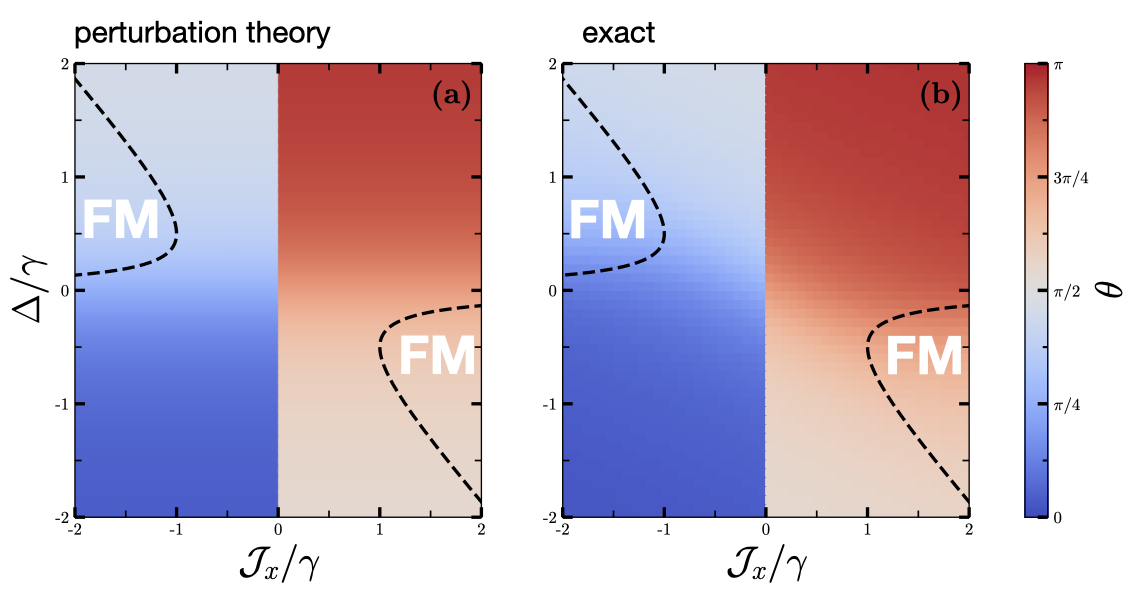}
  \caption{Optimal angle $\theta$ in the $(x,y)$-plain that minimises the spin squeezing parameter in the TFI model as predicted by (a) the perturbative approach and (b) the exact solution for a system with $N = 20$. Black dashed lines show the mean-field phase transition between the paramagnetic phase and ferromagnetic phase.
  }
  \label{fig:TFIphase}
\end{figure*}

\begin{figure}[ht!]
\begin{center}
\includegraphics[width=0.9\columnwidth]{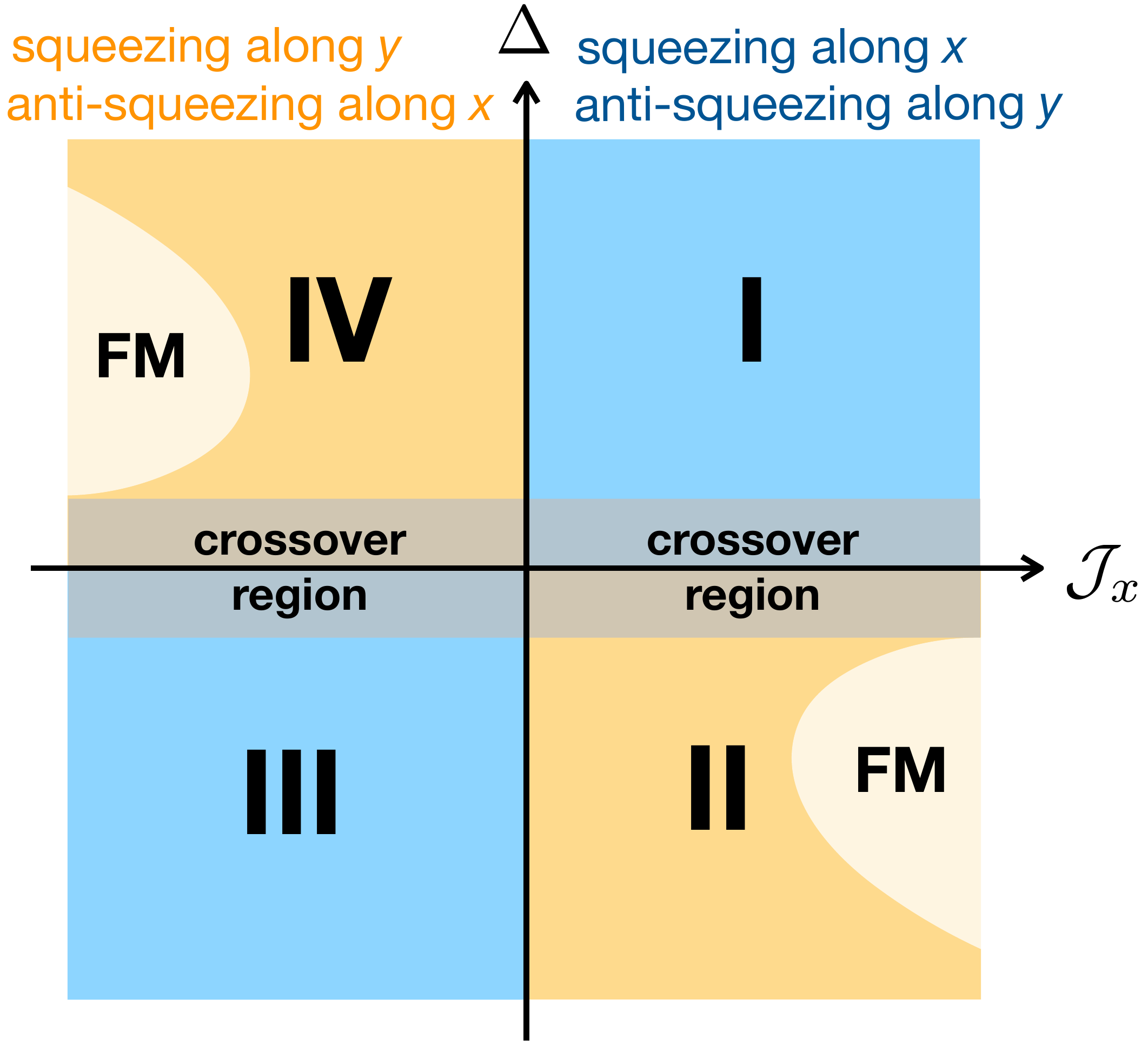}
\caption{Sketch of the steady-state phase diagram of the dissipative TFI model, indicating the squeezed/anti-squeezed collective-spin components predicted by perturbation theory.}
\label{fig:PhDsketch_TFI}
\end{center}
\end{figure}

The matrix element in the numerator of the first-order correction to the steady state reads instead
\begin{align}
\langle \phi_2 | H_1 | \phi_0 \rangle & =   \frac{{\cal J}_x}{4N} \left\langle \phi_2 \right\vert(J^+)^2\left\vert\phi_0  \right\rangle, \\
& =  \frac{{\cal J}_x}{4} \sqrt{\frac{2\left(N-1\right)}{N}}~.
\end{align}
As a consequence the $F$ function entering in the squeezing parameter reads
\begin{align}
F(\{\theta_i\}) = \frac{e^{i2\theta}}{8} \frac{{\cal J}_x(N-1)}{2\Delta - i\gamma} = \frac{\alpha e^{i2\theta}}{\beta-i\gamma},
\end{align}
where $\alpha$ and $\beta$ are given by
\begin{align}\label{eq:alphaTFI}
    \alpha & =\frac{1}{8}{\cal J}_x (N-1) \\
\label{eq:betaTFI}
    \beta & = 2\Delta~.
\end{align}
The $F$ function has the same form as for the XYZ model, Eq.~\eqref{eq:complexreprF}, and therefore the angle $\theta$ extremizing it and the corresponding extremal value have the same expression as those in Eqs.~\eqref{eq:theta_ex} and \eqref{eq:ReFextremalvalue}. As a consequence, for any (finite) value of the perturbation ${\cal J}_x$ and of the field $\Delta$ one can always find an angle $\theta$ for which squeezing appears. 
Fig. \ref{fig:tfi} shows an example of the optimal squeezing parameter and optimal angle for a system with $N = 20$ spins, and for a field with strength $\Delta = -6\gamma$. The pure-state perturbation results capture the linear onset of squeezing upon adding a finite ${\cal J}_x$ perturbation. The agreement with the exact result -- obtained again with the technique developed in Ref.~\cite{ShammahPRA98} -- degrades upon increasing the perturbation because the steady state develops some finite entropy, and most importantly because the system can experience a phase transition to ferromagnetism \cite{OverbeckPRA17} (see Fig.~\ref{fig:PhDsketch_TFI} for a sketch of the phase diagram).

The full dependence of the squeezing angle on the Hamiltonian parameters ${\cal J}_x$ and $\Delta$, and on the dissipation strength $\gamma$, is shown in Fig.~\ref{fig:TFIphase}. To understand the main features of the angle map across the phase diagram, we can extract the optimal squeezing angle simply by taking the limit of negligible dissipation ($|\gamma/\beta| \ll 1$) in Eq.~\eqref{eq:ReF}, which gives ${\rm Re}F \approx \alpha \cos(2\theta) / \beta$. This implies that $\theta=0 ~({\rm mod}~\pi)$ is the optimal squeezing angle if $\alpha/\beta>0$, i.e. if $\Delta>0$ and ${\cal J}_x>0$, or $\Delta<0$ and ${\cal J}_x<0$ (quadrants I and III in the $({\cal J}_x,\Delta)$ plane -- see Fig.~\ref{fig:PhDsketch_TFI}); while $\theta=\pi/2 ~({\rm mod}~\pi)$ is the optimal squeezing angle in the two complementary quadrants $\Delta>0$ and ${\cal J}_x<0$, or $\Delta<0$ and ${\cal J}_x>0$ (quadrants II and IV respectively). The presence of a finite dissipation $\gamma$ introduces a smooth crossover in the squeezing angle between quadrant I and quadrant II, and between quadrant III and quadrant IV.
Fig.~\ref{fig:PhDsketch_TFI} summarizes the structure of correlations (i.e. the squeezed vs. anti-squeezed collective-spin component) across the phase diagram of the system. 

\section{Single-emitter drive and second order perturbation theory}
\label{sec:drivenDicke}

 In this section we consider the most elementary form of drive which perturbs a dissipative system of emitters out of the state with all emitters in the ground state -- namely single-emitter (or Rabi) drive. This generically corresponds to a perturbation in the form 
\begin{equation}
\lambda {H}_1 = - \frac{1}{2} \sum_i \Omega_i \left(e^{i\phi_i} S_i^+ +  e^{-i\phi_i}  S_i^- \right)
\label{e.single_emitter}
\end{equation}
where $\Omega_i$ is the local amplitude of the Rabi drive at site $i$, and $\phi_i$ the local phase. 

The state perturbed by $H_1$ to first order experiences a net rotation of the collective spin away from the $z$ axis, but it remains a coherent spin state without any form of correlations. Indeed the first-order perturbed state is a superposition between the unperturbed state $|\phi_0\rangle \in {\cal S}_{-N/2}$ and a state in the magnetization sector with $M=-N/2+1$, 
$|\phi_1 \rangle \in {\cal S}_{-N/2+1}$, so that $ \langle \phi_1 | S_i^+ S_j^- | \phi_0 \rangle =  \langle \phi_1 | S_i^+ S_j^+ | \phi_0 \rangle = 0$ for any $i \neq j$. 
To observe correlations one needs the superposition of $|\phi_0\rangle$ with a state $|\phi_2\rangle$ in the magnetization sector ${\cal S}_{-M/2+2}$, 
which can only be achieved at second order in the perturbation. Moreover the appearance of correlations requires also an interacting unperturbed Hamiltonian and/or collective emission, given that otherwise the state of the emitters remains perfectly factorized as in the unperturbed case.  

In the following we apply the results of Sec.~\ref{sec:perttheory} for second-order perturbation theory to the relevant example of the driven Dicke model. 

\subsection{Example: the driven Dicke model}\label{subsec:example_driven_DM}

The driven Dicke model \cite{Narduccietal1978,HannukainenL2018} describes a symmetric coupling between the emitters and their environment, with a jump operator $\sqrt{\Gamma/N} J^{-}$. The emitters are driven away from their ground state by a uniform Rabi field along \emph{e.g.} the $x$-direction, namely the unperturbed Hamiltonian is vanishing, $H_0=0$, while the perturbing Hamiltonian simply reads:
\begin{equation}
    \lambda H_1 = -\Omega J^x~
\end{equation}
namely it is of the form of Eq.~\eqref{e.single_emitter} with uniform $\Omega_i =\Omega$ and $\phi_i = 0$. 

As already pointed out above, within first-order perturbation theory only the average collective spin can rotate, but correlations cannot appear; while they appear instead to second order. The pure-state assumption might appear to be unjustified to second order in perturbation theory, since, as discussed in Sec.~\ref{sec:pure}, the steady state of the system generally loses purity to second order in the perturbation.  Nonetheless, the driven Dicke model is a special case, since, as already shown in Ref.~\cite{HannukainenL2018}, and as we shall further exhibit numerically in App.~\ref{sec:DDMpurity}, the purity of the steady state converges exponentially to unity with system size in the normal phase. This justifies again the pure-state assumption behind our approach.

Applying Eqs.~\eqref{e.psi1} and \eqref{e.psi2} we obtain for the perturbed state the form:
\begin{align}
& |\phi_0' \rangle =  \left ( 1-\frac{N\Omega^2}{2\Gamma^2} \right ) \left | -\frac{N}{2} \right \rangle  + \frac{i\Omega \sqrt{N}}{\Gamma }  \left | -\frac{N}{2} +1 \right \rangle  \nonumber \\
& - \frac{\Omega^2 N^2}{\Gamma^2 \sqrt{2N(N-1)}}   \left | -\frac{N}{2} +2 \right \rangle + \mathcal{O}\left ( \frac{\Omega^3}{\Gamma^3} \right )
\label{e.perturbedstate}
\end{align}
where we have adopted the shorthand notation $|J=N/2,M\rangle \to |M\rangle$ for Dicke states. Expressions for various expectation values with respect to $\vert\phi'_0\rangle$ can be found in App.~\ref{app:appB}. We would like to point out that the collective emission plays a crucial role in keeping the perturbed state in the symmetric Dicke manifold, thereby allowing for the appearance of correlations, as the Dicke states are generically entangled except for the fully polarized ones. If the emission were not fully collective, the last term in Eq.~\eqref{e.psi2} would admix the unperturbed state with states going out of the symmetric manifold. 

The perturbed state develops a rotated net collective spin 
\begin{equation}
\langle {\bm J} \rangle = \frac{N}{2} (0, \cos \phi, \sin\phi) = \frac{N}{2} \bm e_{\phi}
\end{equation}
where 
\begin{align}
\cos\phi &  = -\frac{2\Omega}{\Gamma}  + \mathcal{O}\left ( \frac{\Omega^3}{\Gamma^3} \right ) \nonumber \\
 \sin\phi & = - \left ( 1 - \frac{2\Omega^2}{ \Gamma^2} \right ) + \mathcal{O}\left ( \frac{\Omega^3}{\Gamma^3} \right )~.
\end{align}

The squeezed component of the collective spin must therefore be searched for in the plane transverse to this rotated collective spin, namely as 
\begin{equation}
J^\theta = \cos\theta~ J^x + \sin\theta J^{\phi'}~
\end{equation}
where $J^{\phi'} =  - \sin\phi ~J^y + \cos\phi ~J^z$ is the collective spin component orthogonal to $\bm e_\phi$ in the $yz$ plane. 

By construction $J^\theta$ is orthogonal to the net orientation of the collective spin, and therefore $\langle J^\theta \rangle = \langle J^x \rangle = \langle J^{\phi'} \rangle= 0$.  
On the other hand 
\begin{align}
\langle (J^\theta)^2 \rangle & =  \cos^2\theta ~ \langle (J^x)^2 \rangle   + \sin^2\theta ~ \langle (J^{\phi'})^2 \rangle \nonumber \\
 & + \cos\theta \sin\theta ~\langle J^x J^{\phi'} + J^{\phi'} J^x \rangle ~.
\end{align}
One finds that $\langle J^x J^{\phi'} + J^{\phi'} J^x \rangle = 0$ for the perturbed state in Eq.~\eqref{e.perturbedstate}. To evaluate the rest of the above expression, after a lengthy but straightforward calculation one can show that  
\begin{equation}
 \langle (J^x)^2 \rangle = \frac{N}{4} - \frac{N\Omega^2}{2\Gamma^2} +\mathcal{O}\left ( \frac{\Omega^3}{\Gamma^3} \right )
\end{equation}
and
\begin{equation}
 \langle ( J^{\phi '} ) ^2 \rangle = \frac{N}{4} + \frac{N\Omega^2}{2\Gamma^2} + \mathcal{O}\left ( \frac{\Omega^3}{\Gamma^3} \right )~.
\end{equation}
Hence we find that the  $J^x$ fluctuations are squeezed below the projection noise $N/4$, while those of $J^{\phi '}$ are anti-squeezed. 
In fact these collective spin components correspond to those maximizing (resp. minimizing) the squeezing parameter, since
\begin{align}
\xi_R^2 & =   \frac{N~ {\rm Var}(J^\theta)}{\langle \bm J \rangle^2  } \nonumber \\
& = 1 + \frac{2\Omega^2}{\Gamma^2} \left (\sin^2 \theta - \cos^2\theta \right ) 
\end{align}
which is clearly minimized for $\theta = 0$, namely the maximally squeezed spin component corresponds to $J^x$. 
Moreover, we observe that 
\begin{equation}
\langle (J^x)^2 \rangle  \langle (J^{\phi '})^2 \rangle = \frac{|\langle \bm J \rangle|^2}{4} + \mathcal{O}\left ( \frac{\Omega^3}{\Gamma^3} \right )
\end{equation}
namely the squeezed state is again of minimal uncertainty. 

\begin{figure}[t]
  \centering
  \includegraphics[width=0.5\textwidth]{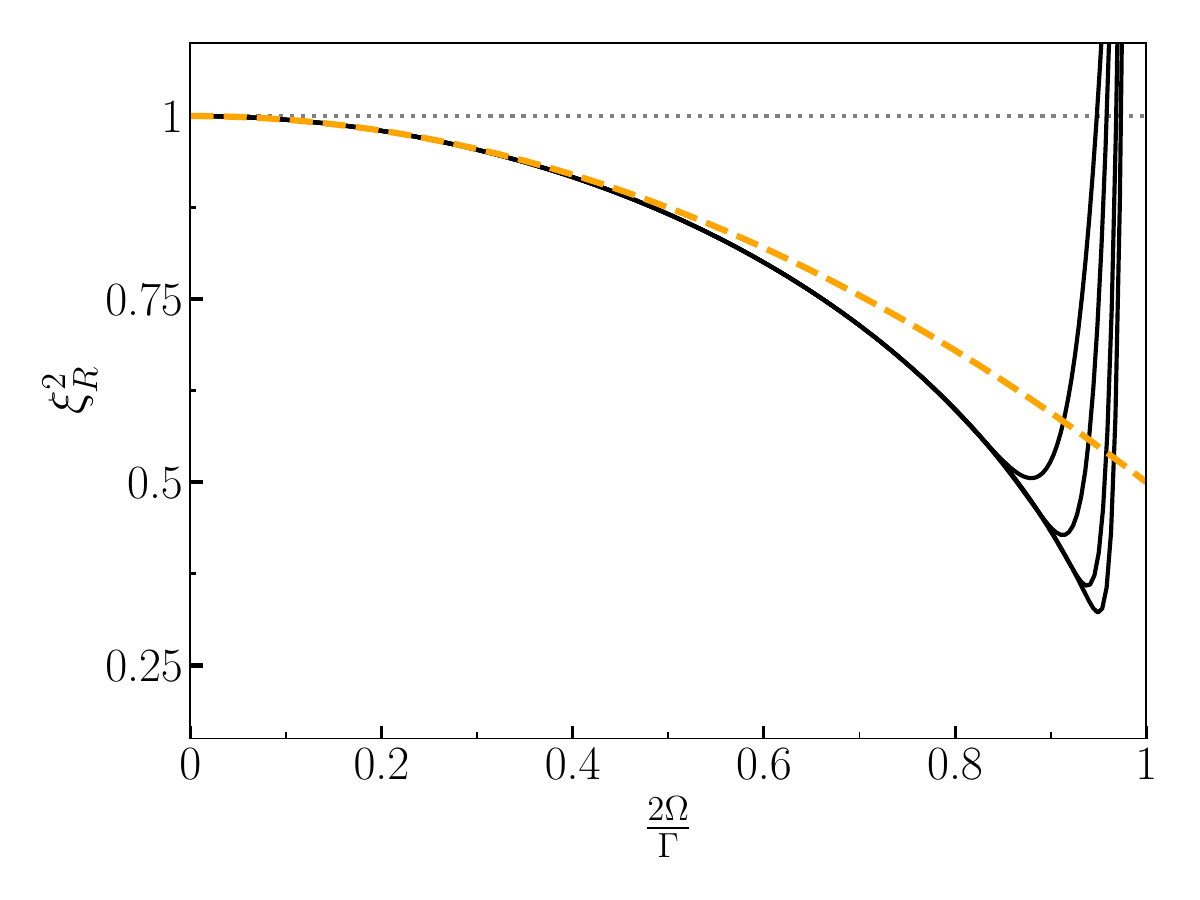}
  \caption{Minimal spin squeezing parameter $\xi_{\rm R}^2$ in the dissipative TFI model as a function of $\frac{2\Omega}{\Gamma}$. The black lines show the exact solutions for $N = 50, 100, 200, 300$ (decreasing minimum as $N$ increases). The orange dashed line shows the result of the second order perturbative approach.
  }
  \label{fig:ddm}
\end{figure}

The results for the minimal spin squeezing parameter ($\theta = 0$) from second-order perturbation theory are compared to the exact results in Fig. \ref{fig:ddm}. The second-order approach reproduces the quadratic onset of squeezing as a function of the drive $\Omega$, and follows its quadratic development for a rather broad parameter range -- roughly equaling half of the extension of the paramagnetic phase in the steady state. Second order perturbation theory clearly breaks down on approach to the super-radiant transition \cite{Narduccietal1978,HannukainenL2018}, occurring for $\Omega/\Gamma = 1/2$.

\section{Discussion}
\label{sec:discussion}

Throughout this work we have shown that spin squeezing is a form of quantum correlations which is promoted by the interplay between Hamiltonian driving and dissipation, either in the form of individual emission or of collective emission. A few previous works already identified squeezing as a characteristic feature of steady states in driven-dissipative quantum systems. In particular Ref.~\cite{LeePRL13} found squeezing in the paramagnetic phase of the dissipative XYZ model on two- and three-dimensional lattices and nearest-neighbor interactions by evaluating Gaussian quantum fluctuations around the mean-field steady state. This result has been confirmed in Ref.~\cite{Verstraelenetal2023} for the two-dimensional dissipative XYZ model, by making use of a truncated cumulant expansion of quantum fluctuations for each pure state within a quantum-trajectory approach to the driven dissipative dynamics. 
Spin squeezing in the paramagnetic steady state of the driven Dicke model has been previously reported in Ref.~\cite{HannukainenL2018}. 
Our work encompasses all these results for weak driving, by defining the general conditions under which squeezing is expected in the paramagnetic phase, close to  the trivial (i.e. undriven) steady state.

All of the applications we discussed in this work focus on collective-spin models (namely on Hamiltonians $H_0$ and $H_1$ only depending on the collective spin operator ${\bm J}$); yet the results we presented arguably apply to a much broader class of models. As discussed in Sec.~\ref{s.extremal}, first- and second-order perturbation theory starting from the coherent spin state  $|\phi_0\rangle$ can only produce states with $J\geq N/2-2$. Hence both the unperturbed Hamiltonian $H_0$ as well as its perturbation $H_1$ act on the states of interest as projected onto this large-spin sector of Hilbert space 
\begin{equation}
H_{0(1)} \to H^{\rm (eff)}_{0(1)} = {\cal P}_{J\geq N/2-2}  ~ H_{0(1)}  ~ {\cal P}_{J\geq N/2-2} 
\end{equation}
where ${\cal P}_{J\geq N/2-2}$ is the projector on the sector with $J\geq N/2-2$. Regardless of the range of interactions in $H_{0(1)}$, the projected Hamiltonians $H^{\rm (eff)}_{0(1)}$ have effectively long-range interactions. As an example, the Hamiltonian projected onto the sector with maximal $J=N/2$ has only infinite-range interactions (namely it depends only on the ${\bm J}$ operator). Therefore the physics of the collective-spin models explored in this work applies in fact also to their short-range-interacting versions, as long as one focuses on perturbative effects close to the $U(1)$ symmetric limit.

Single-emitter driving as in the driven Dicke model is obviously very relevant to experiments, as it is obtained by coupling the emitters to a Rabi field. The physics of the driven Dicke model has been recently realized experimentally in Ref.~\cite{Ferioli21}, and one can therefore expect squeezing to be a feature of the steady state at weak driving, and possibly up to the super-radiant transition. On the other hand, the experimental relevance of two-emitter driving is less obvious. Nonetheless, as discussed in Ref.~\cite{LeePRL13}, two-emitter driving can be realized by coupling pairs of emitters to pairs of lasers that are strongly detuned from single-emitter transitions, but are instead resonant for two-photon two-emitter transitions.  

\section{Conclusions}
\label{sec:conclusions}

In this work we have introduced pure-state perturbation theory for the steady state of open quantum systems, valid when the system Hamiltonian is perturbed away from a known limit admitting a pure state as steady state. The assumption of a pure perturbed state allows for closed expressions of the coefficient of the perturbed state on the unperturbed Hamiltonian eigenbasis, and particularly so when the non-Hermitian Hamiltonian of the open quantum system is diagonalized by the eigenbasis of the unperturbed Hamiltonian. 

Pure-state perturbation theory offers a significant insight into the physical properties of the perturbed steady state, in particular for what concerns quantum correlations. In this work we have focused our attention on ensembles of light emitters -- i.e. qubits or (artificial) atoms -- and we characterized quantum correlations by considering collective-spin properties, in particular the uncertainty of its components perpendicular to the average orientation. For ensembles of light emitters which emit photons individually, we could formulate a theorem for the onset of spin squeezing under the effect of two-emitter driving, based on first-order perturbation theory. The theorem applies to relevant models of collective interactions among emitters, i.e. the dissipative XYZ model \cite{HuybrechtsPRB20} and the dissipative transverse-field Ising model \cite{OverbeckPRA17}  with all-to-all couplings. Furthermore we showed that single-emitter drive, in the form of Rabi coupling, leads to spin squeezing when the emitters are collectively coupled to the environment, as modeled by the driven Dicke model. 
Perturbation theory allows for explicit expressions of the perturbed states, and it shows that the spin-squeezed states are of minimal uncertainty, namely  spin squeezing is their optimal resource for metrology, since the inverse spin-squeezing parameter coincides with the quantum Fisher information. 

Our pure-state perturbation results are tested against exact results, and they exhibit a rather large domain of accuracy away from the unperturbed limit, validating the pure-state assumption. The near purity, or low entropy, of the perturbed state allows the Hamiltonian perturbation to induce \emph{quantum} correlations in the steady state. It is important to point out that the low entropy of the steady state is actually a condition valid when the dissipation is much stronger than the drive, keeping the system close to the (pure) ground state of each emitter. The fast dissipation prevents the emitters from being highly excited and therefore from radiating significantly into the environment -- something which would entail instead a significant entropy in the steady state. Hence the form of quantum correlations described in this work are the result of a very specific interplay between dissipation and drive, resulting in the so-called normal phase within the steady-state phase diagram. This can be contrasted with the super-radiant phase, developing long-range correlations among the emitters, but also a large entropy due to the important emission of radiation into the environment. 

The perturbative approach allows for analytical predictions of the steady state and its quantum correlation properties when the system is weakly driven. But, more generally, the predictions of perturbation theory provide a significant insight into the mechanisms leading to the appearance of the specific correlation patterns of the steady state, which can characterize the phase diagram beyond the strict limit of applicability of the perturbative approach. This is crucial in the absence of a simple variational principle -- such as the minimization of the free energy -- which dictates the form of the steady state, and which could allow for an intuitive (\emph{e.g.} semiclassical) prediction of its properties. A relevant example provided in this work is offered by the appearance of ferromagnetic correlations in the steady state of the dissipative XYZ model, in spite of the Hamiltonian interactions being antiferromagnetic; and the fact  that the most correlated spin components are not necessarily those which are most strongly interacting. These predictions go clearly beyond what one can reconstruct via mean-field theory, which pictures the whole paramagnetic phase as being identical to the U(1)-symmetric steady state \cite{HuybrechtsPRB20}. The insight acquired via perturbation theory can be used as a precious guidance in the understanding and the design of many-body quantum states resulting from the competition between strong dissipation and a weak drive; and it can be applied rather broadly to dissipative quantum many-body systems. In this work we focused on ensembles of light emitters, relevant to a very broad spectrum of physical systems ranging from atomic physics (e.g. Rydberg atoms \cite{LeePRL13}, atomic ensembles close to the super-radiant transition \cite{HannukainenL2018, Ferioli21}, etc.) to solid-state physics \cite{Cong16}. Yet it is clear that one can apply the same scheme to different degrees of freedom -- \emph{e.g.} to coupled bosonic modes leaking into the environment, and close to their vacuum state. 

\section{Acknowledgements}
Useful discussions with I. Ferrier-Barbut and F. Minganti are gratefully acknowledged. This work is supported by QuantERA (MAQS project) and Agence National de la Recherche (ANR-25-CE57-1786, project STEFAN, and ANR-22-PETQ-0004 France 2030, project QuBitAF).

%\bibliography{biblio.bib}
\input{main.bbl}

\newpage
\onecolumn

\newpage
\appendix
\onecolumn

\section{Proof of the purity of the perturbed state to first order}\label{sec:proof}

In this section we provide a formal proof of the general fact that, perturbing a pure steady state, one obtains again a pure state to first order in perturbation theory. 

We consider a Liouvillian of the form $\mathcal{L} = \mathcal{L}_0 + \lambda \mathcal{L}_1$, with $\mathcal{L}_0$ an unperturbed Liouvillian with a pure steady state $\rho^{(0)}_s = \vert\phi_0\rangle\langle\phi_0\vert$, and $\mathcal{L}_1$ a generic perturbation such that $\mathcal{L}$ is still a Liouvillian. One can, without loss of generality, rewrite the density matrix in a basis $\{ |\phi_n \rangle \}$ where the pure steady state is a basis element, so that it reads:
\begin{equation}
    \rho_s^{(0)} = 
\begin{pmatrix}
    0 &\dots & 0 \\
    \vdots & \ddots & \vdots \\
    0 & \dots & 1
\end{pmatrix}~.
\end{equation}
One can write the first-order correction $\rho_s^{(1)}$ as (see Ref. \cite{Li14})
\begin{equation}\label{eq:pertact}
\rho_s^{(1)}=-\mathcal{L}_0^{-1}\left[\mathcal{L}_1\left[\rho_s^{(0)}\right]\right] = \begin{pmatrix}
c_{00} & c_{01} & \dots &  c_{0,D-1}\\
c_{10}&  c_{11} & \dots  & c_{1,D-1}\\
 \vdots&\vdots & \ddots & \vdots\\
c_{D-1,0}&  c_{D-1,1} & \dots&  c_{D-1,D-1}
\end{pmatrix},
\end{equation}
with $\mathcal{L}_0^{-1}$ the (pseudo)inverse of $\mathcal{L}_0$ (see below). Here $D$ is the Hilbert-space dimension. It follows that the purity of the perturbed state to first order, $\rho_s = \rho_s^{(0)} + \lambda \rho_s^{(1)} + {\cal O}(\lambda^2) = \vert\phi_0\rangle\langle\phi_0\vert + \lambda\sum_{nm} c_{nm}\vert\phi_n\rangle\langle\phi_m\vert  + {\cal O}(\lambda^2)$, reads
\begin{equation}\label{eq:purity}
    {\rm Tr}[\rho_s^2] = 1 + 2~\lambda~ c_{00} + {\cal O}(\lambda^2)~.
\end{equation}
It thus suffices to show that $c_{00}=\text{Tr}[\rho_s^{(0)}\rho_s^{(1)}]=0$ for the purity to be at most quadratic in the perturbation. \

To show this, we turn our attention to the matrix properties of the unperturbed Liouvillian and its (pseudo)inverse, by writing them in a superoperator formalism. The elements of the Liouvillian superoperator (i.e. an operator transforming operators into operators) can be formally written as 
\begin{equation}
    \hat{\mathcal{L}}_0 = \sum_{n,m,k,l} a_{nmkl}\vert n\rangle\vert m\rangle\langle k\vert\langle l\vert,
\end{equation}
where $\vert n\rangle\vert m\rangle = \vert n\rangle\otimes\vert m\rangle$, and the hat notation denotes a superoperator form.  Note that the eigenvectors of this superoperator are vectorised operators, based on the Choi-Jamiolkowski isomorphism $\vert n\rangle\langle m\vert \leftrightarrow \vert n\rangle\otimes\vert m \rangle = |nm\rangle$.

Any density matrix $\rho$ can be written in its vectorised form $\vert {\rho}\rangle$. Hence for the steady state one can write
\begin{equation}
    \hat{\mathcal{L}}_0\vert \rho_s^{(0)}\rangle = \lambda_0 \vert\rho_s^{(0)}\rangle = 0~.
\end{equation}
This vanishing eigenvalue makes the unperturbed Liouvillian non-invertible, justifying the need to resort to pseudo-inversion. We choose here Moore-Penrose pseudo-inversion, which can be formulated via a singular-value decomposition (SVD) of the Liouvillian. The SVD reads
\begin{equation}
    \hat{\mathcal{L}}_0 = U\Sigma V^\dagger, 
\end{equation}
and its Moore-Penrose pseudo-inverse, denoted as $\hat{\mathcal{L}}_0^\neg$, is defined as follows:
\begin{equation}
    \hat{\mathcal{L}}_0^{\neg} = V\Sigma^\neg U^\dagger,
\end{equation}
where $\Sigma^\neg$ can be calculated from the matrix $\Sigma$ with singular values $s_i$ on its diagonal, by replacing $s_i$ with $1/s_i$ when $s_i\neq 0$, and leaving $s_i$ as such when $s_i = 0$. The presence of a zero eigenvalue of the Liouvillian, associated with the steady state, implies a zero singular value $s_0 = 0$. 
At this stage,  it is useful to recall the relation between the SVD and the eigenvalue decomposition. Given a matrix $M$ with SVD $M=U\Sigma V^\dagger$,  it follows that 
\begin{equation}
    M^\dagger M = V\Sigma^\dagger U^\dagger U \Sigma V^\dagger = V (\Sigma^\dagger\Sigma) V^\dagger,
\end{equation}
more specifically, the right-hand side of the equation corresponds to the eigenvalue decomposition of the left-hand side. It follows that (i) the singular values of $\Sigma$ correspond to the square root of the eigenvalues of $M^\dagger M$, and (ii) the columns of $V$ are the eigenvectors of $M^\dagger M$. In our problem of interest the eigenvector corresponding to the eigenvalue zero is unique, and it corresponds to the unperturbed steady state $|\phi_0\rangle |\phi_0\rangle$. Since $\vert\phi_0\rangle\vert\phi_0\rangle$ is an eigenvector of $\mathcal{L}_0$ with eigenvalue zero, the same is true for $\mathcal{L}_0^\dag\mathcal{L}_0$. Hence $|\phi_0\rangle |\phi_0\rangle$ is a column of $V$, and all other columns are orthogonal to it due to the unitarity of $V$. Representing $V$ in the basis of eigenstates of ${\cal L}_0$, it can therefore be written, together with $\Sigma$, as 
\begin{equation}
    V = 
    \begin{pmatrix}
     1 & 0 & \dots & 0 \\
      0 &  v_{11} & \dots & v_{1,D^2-1}\\
      \vdots & \vdots & \ddots & \vdots \\
      0 &  v_{D^2-1,1} & \dots &  v_{D^2-1,D^2-1} 
    \end{pmatrix}~~~~
    \Sigma = \begin{pmatrix}
          0 & 0 & 0 & 0 \\
        0 & s_1 & \dots &  0 \\
        \vdots & \vdots  & \ddots & 0 \\
        0 & 0 & 0  & s_{D^2-1} 
    \end{pmatrix}~.
    \quad
\end{equation}

Without any assumption on the specific form of $U$, we can use the above expressions to calculate $\hat{\mathcal{L}}_0^{\neg}=V\Sigma^\neg U^\dagger$, where

\begin{equation}
\Sigma^{\neg} = \begin{pmatrix}
          0 & 0 & 0 & 0 \\
        0 & 1/s_1 & \dots &  0 \\
        \vdots & \vdots  & \ddots & 0 \\
        0 & 0 & 0  & 1/s_{D^2-1} 
    \end{pmatrix}~.
\end{equation}

It is easy to realize that $\hat{\mathcal{L}}_0^{\neg}$ has a vanishing first row
\begin{equation}
\hat{\mathcal{L}}_0^\neg = 
\begin{pmatrix}
0& 0 & \dots & 0 \\
l_{10} & l_{11} & \dots &   l_{0, D^2-1}\\
 \vdots&\vdots & \ddots & \vdots\\
l_{D^2-1,0}& l_{D^2-1,1} & \dots & l_{D^2-1,D^2-1}
\end{pmatrix},
\end{equation}
and, as a consequence, when acting upon any vector, it produces a vector \emph{orthogonal} to the $|\phi_0\rangle|\phi_0\rangle$ vector
\begin{equation}
 \hat{\mathcal{L}}_0^\neg \vert A \rangle
=
\begin{pmatrix}
 0 \\
 A_1' \\
 \dots \\
A'_{D^2-1}
\end{pmatrix}~.
\end{equation}

Hence this proves that $\rho_s^{(1)}$ in Eq.~\eqref{eq:pertact} is orthogonal to $\rho_s^{(0)}$, leading to the result $c_{00}=0$, namely $\text{Tr}[\rho_s^2] = 1 + \mathcal{O}(\lambda^2)$. 

\begin{figure}[h]
  \centering
  \includegraphics[width=0.7\textwidth]{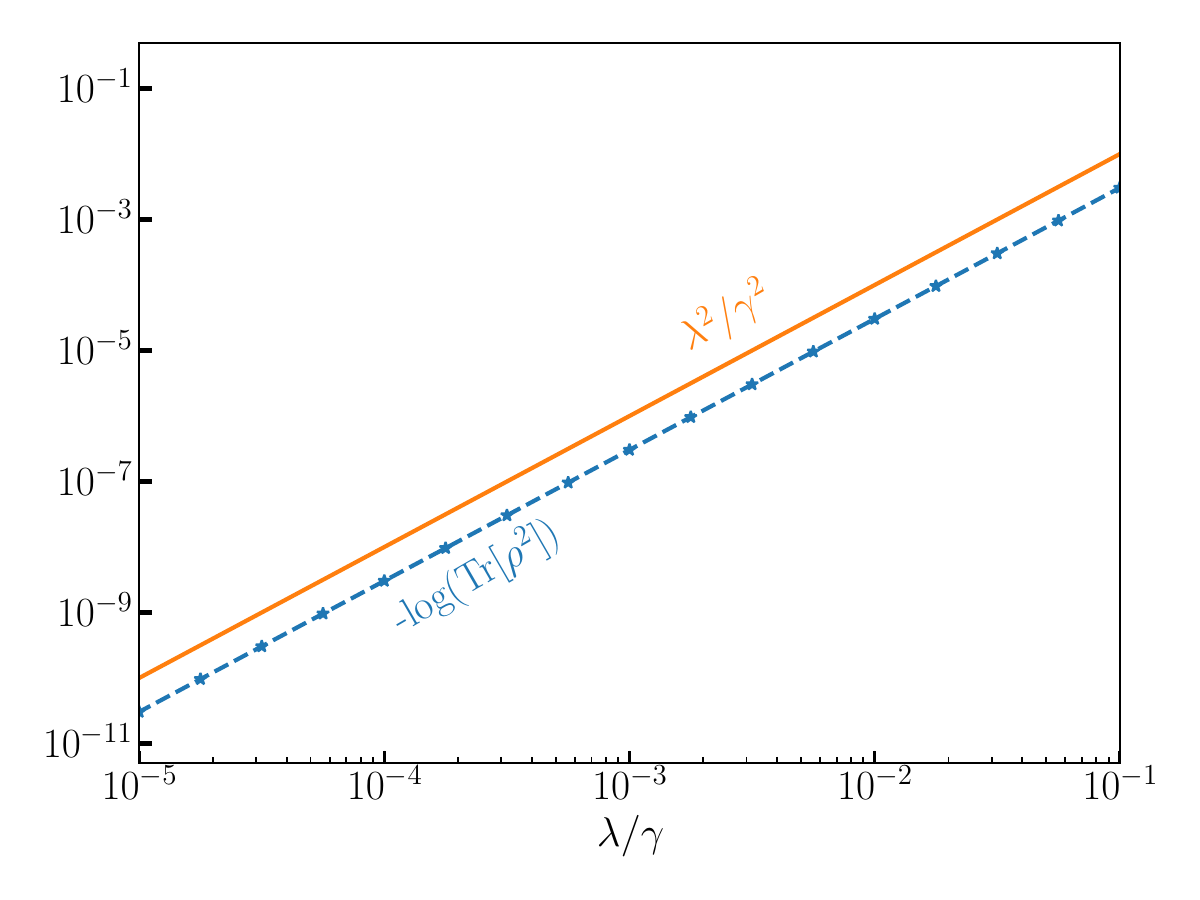}
  \caption{Second R\'enyi entropy (blue star markers) for the dissipative XYZ model from Sec.~\ref{sec:xyz} for $N=90$ spins, $\mathcal{J}=-0.2\gamma$, $\mathcal{J}_z = \gamma$ and $\lambda = \delta\mathcal{J}$. The orange line denotes the behavior $\sim \lambda^2$.}
  \label{fig:scaling}
\end{figure}

Fig.~\ref{fig:scaling} shows the second R\'enyi entropy $S_2 = - \log[{\rm Tr}(\rho_s^2)]$ of the exact steady state for the dissipative XYZ model as a function of the perturbation. The quadratic dependence on $\lambda$ is clearly manifested, confirming that the $\lambda^2$ term is the leading-order perturbative correction to the purity.

\section{Calculation of ${\rm Re}\left[F(\left\{\theta_{\rm ex}\right\})\right]$}\label{app:ReFex}
In this section we detail the calculations of the function ${\rm Re}\left[F(\left\{\theta_{\rm ex}\right\})\right]$ appearing in theorem \ref{theorem1}, assuming that its real part can be written in the form of Eq.~\eqref{eq:ReF}. That is, one can write 
\begin{equation}\label{eqapp:F}
{\rm Re}\left[F(\left\{\theta_{\rm ex}\right\})\right] = \frac{\alpha\left(\beta\cos2\theta_{\rm ex} - \gamma\sin2\theta_{\rm ex}\right)}{\beta^2 + \gamma^2}.
\end{equation}
The angle $\theta_{\rm ex}$ extremizing the function leads to a vanishing derivative of Eq.~\eqref{eqapp:F} with respect to $\theta$ to zero. The result of Eq.~\eqref{eq:theta_ex} is then obtained as follows
\begin{align}
\frac{d}{d\theta}{\rm Re}[F(\{\theta\})] = 0 ~~~~
\Leftrightarrow ~~~~ \tan 2\theta  =  -\frac{\gamma}{\beta} ~~~~
\Leftrightarrow ~~~~ \theta_{ex} = \frac{1}{2}\tan^{-1}\left(-\frac{\gamma}{\beta}\right) + k\frac{\pi}{2} \quad k\in\mathbb{Z}.
\end{align}
We can now substitute this result in Eq.~\eqref{eqapp:F}, and calculate for which angle ${\rm Re}\left[F(\left\{\theta_{\rm ex}\right\})\right] > 0$, i.e. spin squeezing occurs. This yields
\begin{equation}\label{eqapp:ReFcalc}
    {\rm Re}\left[F(\left\{\theta_{\rm ex}\right\})\right] = \frac{\alpha\left[\beta\cos\left(\tan^{-1}\left(-\frac{\gamma}{\beta}\right) + k\pi \right) - \gamma\sin\left(\tan^{-1}\left(-\frac{\gamma}{\beta}\right) + k\pi \right)\right]}{\beta^2 + \gamma^2}~.
\end{equation}
We can simplify this expression via standard trigonometric formulas, and using the fact that  $\cos(\tan^{-1}(x)) = \frac{1}{\sqrt{1+x^2}}$ and $\sin(\tan^{-1}(x)) = \frac{x}{\sqrt{1+x^2}}$. Hence,
\begin{align}
    \cos\left(\tan^{-1}\left(-\frac{\gamma}{\beta}\right) + k\pi \right) 
     = \frac{\cos(k\pi)}{\sqrt{1 + \frac{\gamma^2}{\beta^2}}},
\end{align}
and
\begin{align}
    \sin\left(\tan^{-1}\left(-\frac{\gamma}{\beta}\right) + k\pi \right) =  \frac{-\frac{\gamma}{\beta}}{\sqrt{1 + \frac{\gamma^2}{\beta^2}}}\cos\left( k\pi\right).
\end{align}
Substituting this result in Eq.~\eqref{eqapp:ReFcalc} finally allows us to write
\begin{align}
    {\rm Re}\left[F(\left\{\theta_{\rm ex}\right\})\right] = \frac{\alpha\left[\beta^2\frac{\cos(k\pi)}{\sqrt{\beta^2 + \gamma^2}}
    + \gamma^2\frac{\cos\left( k\pi\right)}{\sqrt{\beta^2 + \gamma^2}}
    \right]}{\beta^2 + \gamma^2} = 
\frac{\alpha\cos(k\pi)}{\sqrt{\beta^2 + \gamma^2}}.
\end{align}

\section{Perturbation-theory predictions for the driven Dicke model}\label{app:appB}
In this section we provide the reader with a list of the results for the expectation values of collective spin operators, and products thereof, for the driven Dicke model discussed in Section~\ref{subsec:example_driven_DM}, as obtained for the perturbed state $\vert\phi_0'\rangle$ given in Eq.~\ref{e.perturbedstate}. After a straightforward calculation, one obtains up to second order in $\Omega/\Gamma$:
\begin{align}
    \langle\phi_0'\vert J^x\vert\phi_0'\rangle &= 0~,\\
    \langle\phi_0'\vert J^y\vert\phi_0'\rangle &= -\frac{\Omega N}{\Gamma}~,\\
    \langle\phi_0'\vert J^z\vert\phi_0'\rangle &= -\frac{N}{2} +\frac{\Omega^2N}{\Gamma^2}~, \\ 
    \langle\phi_0'\vert J^z\vert\phi_0'\rangle^2 &= \frac{N^2}{4} - \frac{\Omega^2 N^2}{\Gamma^2 }~,\\
    \langle\phi_0'\vert \left(J^x\right)^2\vert\phi_0'\rangle &=\frac{N}{4}-\frac{\Omega^2 N}{2\Gamma^2}~, \\
    \langle\phi_0'\vert \left(J^y\right)^2\vert\phi_0'\rangle &= \frac{N}{4}+\frac{1}{2}\frac{\Omega^2}{\Gamma^2}N(2N-1)~,\\
    \langle\phi_0'\vert \left(J^z\right)^2\vert\phi_0'\rangle &= \frac{N^2}{4} - \frac{\Omega^2}{\Gamma^2 }N\left(N - 1\right)~.
\end{align}
For the expressions of the form $\langle\phi_0'\vert J^\alpha J^\beta\vert\phi_0'\rangle + \langle\phi_0'\vert J^\beta J^\alpha\vert\phi_0'\rangle$ one finds
\begin{align}
    \langle\phi_0'\vert J^x J^y\vert\phi_0'\rangle + \langle\phi_0'\vert J^y J^x\vert\phi_0'\rangle &= 0~,\\
    \langle\phi_0'\vert J^x J^z\vert\phi_0'\rangle + \langle\phi_0'\vert J^z J^x\vert\phi_0'\rangle &= 0~,\\
    \langle\phi_0'\vert J^y J^z\vert\phi_0'\rangle + \langle\phi_0'\vert J^z J^y\vert\phi_0'\rangle &= \frac{\Omega}{\Gamma}N\left(N-1\right)~.
\end{align}

\section{Purity of the steady state of the driven Dicke model}\label{sec:DDMpurity}
\begin{figure}[h]
  \centering
  \includegraphics[width=1.0\textwidth]{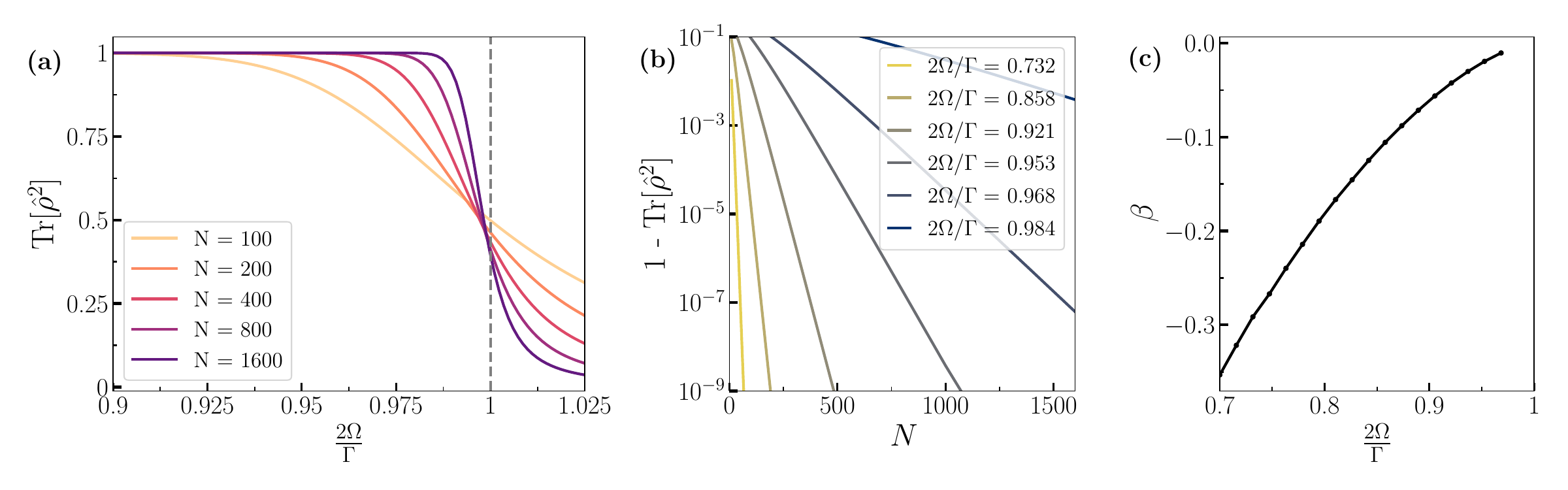}
  \caption{(a) Purity of the steady state of the driven Dicke model as a function of $2\Omega/\Gamma$ for various system sizes; (b) Deviation of the purity from unity as a function of system size for various values of $\frac{2\Omega}{\gamma}$, exhibiting an exponential behavior; (c) Exponent $\beta$ of the fit of the curves in panel (b) to the function $\alpha e^{\beta N}$.}
  \label{fig:purityscaling}
\end{figure}

It has already been observed that the steady state of the driven Dicke model (see Sec.~\ref{subsec:example_driven_DM}) in the normal phase becomes pure in the thermodynamic limit \cite{HannukainenL2018}. Fig.~\ref{fig:purityscaling} presents numerical results illustrating quantitatively this property. Fig.~\ref{fig:purityscaling}(a) shows the purity as a function of $2\Omega/\Gamma$ for various system sizes. In the normal phase ($2\Omega/\Gamma < 1$) the steady state becomes increasingly pure as $N\rightarrow \infty$. In the thermodynamic limit the purity becomes discontinuous at the phase transition, and jumps from a pure state to a maximally mixed state \cite{HannukainenL2018}.
The behavior of the purity as a function of system size for various values of $2\Omega/\Gamma$ is shown in Fig.~\ref{fig:purityscaling}(b). The purity of normal phase is seen to approach unity via an exponential scaling, namely $\text{Tr}[\hat{\rho}^2] = 1 - \alpha e^{\beta N}$. As seen in Fig.~\ref{fig:purityscaling}(c), the absolute value $\vert\beta\vert$ of this exponent decreases as $\frac{2\Omega}{\Gamma}$ approaches the critical value for the normal-to-superradiant transition. This is in line with the results observed in Fig.~\ref{fig:purityscaling}(a), where $\text{Tr}[\hat{\rho}^2]$ is seen to saturate to unity for smaller sizes the smaller $2\Omega/\Gamma$. 

\end{document}

%% file: main.bbl

%% file: main.bbl
\begin{thebibliography}{48}%
\makeatletter
\providecommand \@ifxundefined [1]{%
 \@ifx{#1\undefined}
}%
\providecommand \@ifnum [1]{%
 \ifnum #1\expandafter \@firstoftwo
 \else \expandafter \@secondoftwo
 \fi
}%
\providecommand \@ifx [1]{%
 \ifx #1\expandafter \@firstoftwo
 \else \expandafter \@secondoftwo
 \fi
}%
\providecommand \natexlab [1]{#1}%
\providecommand \enquote  [1]{``#1''}%
\providecommand \bibnamefont  [1]{#1}%
\providecommand \bibfnamefont [1]{#1}%
\providecommand \citenamefont [1]{#1}%
\providecommand \href@noop [0]{\@secondoftwo}%
\providecommand \href [0]{\begingroup \@sanitize@url \@href}%
\providecommand \@href[1]{\@@startlink{#1}\@@href}%
\providecommand \@@href[1]{\endgroup#1\@@endlink}%
\providecommand \@sanitize@url [0]{\catcode `\\12\catcode `\$12\catcode
  `\&12\catcode `\#12\catcode `\^12\catcode `\_12\catcode `\%12\relax}%
\providecommand \@@startlink[1]{}%
\providecommand \@@endlink[0]{}%
\providecommand \url  [0]{\begingroup\@sanitize@url \@url }%
\providecommand \@url [1]{\endgroup\@href {#1}{\urlprefix }}%
\providecommand \urlprefix  [0]{URL }%
\providecommand \Eprint [0]{\href }%
\providecommand \doibase [0]{http://dx.doi.org/}%
\providecommand \selectlanguage [0]{\@gobble}%
\providecommand \bibinfo  [0]{\@secondoftwo}%
\providecommand \bibfield  [0]{\@secondoftwo}%
\providecommand \translation [1]{[#1]}%
\providecommand \BibitemOpen [0]{}%
\providecommand \bibitemStop [0]{}%
\providecommand \bibitemNoStop [0]{.\EOS\space}%
\providecommand \EOS [0]{\spacefactor3000\relax}%
\providecommand \BibitemShut  [1]{\csname bibitem#1\endcsname}%
\let\auto@bib@innerbib\@empty
%</preamble>
\bibitem [{\citenamefont {Fr\'erot}\ \emph {et~al.}(2023)\citenamefont
  {Fr\'erot}, \citenamefont {Fadel},\ and\ \citenamefont
  {Lewenstein}}]{Frerot2023}%
  \BibitemOpen
  \bibinfo {author} {I.~Fr\'erot}, \bibinfo {author} {M.~Fadel}\ and\ \bibinfo
  {author} {M.~Lewenstein},\ \emph {\bibinfo {title} {Probing quantum
  correlations in many-body systems: a review of scalable methods}},\ \href
  {\doibase 10.1088/1361-6633/acf8d7} {\bibfield  {journal} {\bibinfo
  {journal} {Reports on Progress in Physics}\ }\textbf {\bibinfo {volume}
  {86}},\ \bibinfo {pages} {114001} (\bibinfo {year} {2023})}\BibitemShut
  {NoStop}%
\bibitem [{\citenamefont {Horodecki}\ \emph {et~al.}(2009)\citenamefont
  {Horodecki}, \citenamefont {Horodecki}, \citenamefont {Horodecki},\ and\
  \citenamefont {Horodecki}}]{Horodeckietal2009}%
  \BibitemOpen
  \bibinfo {author} {R.~Horodecki}, \bibinfo {author} {P.~Horodecki}, \bibinfo
  {author} {M.~Horodecki}\ and\ \bibinfo {author} {K.~Horodecki},\ \emph
  {\bibinfo {title} {Quantum Entanglement}},\ \href
  {https://doi.org/10.1103/RevModPhys.81.865} {\bibfield  {journal} {\bibinfo
  {journal} {Rev. Mod. Phys.}\ }\textbf {\bibinfo {volume} {81}},\ \bibinfo
  {pages} {865} (\bibinfo {year} {2009})}\BibitemShut {NoStop}%
\bibitem [{\citenamefont {Georgescu}\ \emph {et~al.}(2014)\citenamefont
  {Georgescu}, \citenamefont {Ashhab},\ and\ \citenamefont
  {Nori}}]{Georgescuetal2014}%
  \BibitemOpen
  \bibinfo {author} {I.~M. Georgescu}, \bibinfo {author} {S.~Ashhab}\ and\
  \bibinfo {author} {F.~Nori},\ \emph {\bibinfo {title} {Quantum simulation}},\
  \href {\doibase 10.1103/RevModPhys.86.153} {\bibfield  {journal} {\bibinfo
  {journal} {Rev. Mod. Phys.}\ }\textbf {\bibinfo {volume} {86}},\ \bibinfo
  {pages} {153} (\bibinfo {year} {2014})}\BibitemShut {NoStop}%
\bibitem [{\citenamefont {Verstraete}\ \emph {et~al.}(2009)\citenamefont
  {Verstraete}, \citenamefont {Wolf},\ and\ \citenamefont
  {Cirac}}]{VerstraeteNATPH2009}%
  \BibitemOpen
  \bibinfo {author} {F.~Verstraete}, \bibinfo {author} {M.~M. Wolf}\ and\
  \bibinfo {author} {J.~I. Cirac},\ \emph {\bibinfo {title} {Quantum
  computation and quantum-state engineering driven by dissipation}},\ \href
  {https://doi.org/10.1038/nphys1342} {\bibfield  {journal} {\bibinfo
  {journal} {Nat. Phys.}\ }\textbf {\bibinfo {volume} {5}},\ \bibinfo {pages}
  {633} (\bibinfo {year} {2009})}\BibitemShut {NoStop}%
\bibitem [{\citenamefont {Diehl}\ \emph {et~al.}(2008)\citenamefont {Diehl},
  \citenamefont {Micheli}, \citenamefont {Kantian}, \citenamefont {Kraus},
  \citenamefont {B\"uchler},\ and\ \citenamefont {Zoller}}]{DiehlNATPH2008}%
  \BibitemOpen
  \bibinfo {author} {S.~Diehl}, \bibinfo {author} {A.~Micheli}, \bibinfo
  {author} {A.~Kantian}, \bibinfo {author} {B.~Kraus}, \bibinfo {author} {H.~P.
  B\"uchler}\ and\ \bibinfo {author} {P.~Zoller},\ \emph {\bibinfo {title}
  {Quantum states and phases in driven open quantum systems with cold atoms}},\
  \href {https://doi.org/10.1038/nphys1073} {\bibfield  {journal} {\bibinfo
  {journal} {Nat. Phys.}\ }\textbf {\bibinfo {volume} {4}},\ \bibinfo {pages}
  {878} (\bibinfo {year} {2008})}\BibitemShut {NoStop}%
\bibitem [{\citenamefont {Weimer}\ \emph {et~al.}(2010)\citenamefont {Weimer},
  \citenamefont {M{\"u}ller}, \citenamefont {Lesanovsky}, \citenamefont
  {Zoller},\ and\ \citenamefont {B{\"u}chler}}]{WeimerNPhys2010}%
  \BibitemOpen
  \bibinfo {author} {H.~Weimer}, \bibinfo {author} {M.~M{\"u}ller}, \bibinfo
  {author} {I.~Lesanovsky}, \bibinfo {author} {P.~Zoller}\ and\ \bibinfo
  {author} {H.~P. B{\"u}chler},\ \emph {\bibinfo {title} {A Rydberg quantum
  simulator}},\ \href {\doibase 10.1038/nphys1614} {\bibfield  {journal}
  {\bibinfo  {journal} {Nature Physics}\ }\textbf {\bibinfo {volume} {6}},\
  \bibinfo {pages} {382} (\bibinfo {year} {2010})}\BibitemShut {NoStop}%
\bibitem [{\citenamefont {M\"uller}\ \emph {et~al.}(2012)\citenamefont
  {M\"uller}, \citenamefont {Diehl}, \citenamefont {Pupillo},\ and\
  \citenamefont {Zoller}}]{Mueller_2012}%
  \BibitemOpen
  \bibinfo {author} {M.~M\"uller}, \bibinfo {author} {S.~Diehl}, \bibinfo
  {author} {G.~Pupillo}\ and\ \bibinfo {author} {P.~Zoller},\ \emph {\bibinfo
  {title} {Engineered Open Systems and Quantum Simulations with Atoms and
  Ions}},\ \href {https://doi.org/10.1016/B978-0-12-396482-3.00001-6}
  {\bibfield  {journal} {\bibinfo  {journal} {Adv. At. Mol. Opt. Phys.}\
  }\textbf {\bibinfo {volume} {61}},\ \bibinfo {pages} {1} (\bibinfo {year}
  {2012})}\BibitemShut {NoStop}%
\bibitem [{\citenamefont {Lee}\ \emph {et~al.}(2013)\citenamefont {Lee},
  \citenamefont {Gopalakrishnan},\ and\ \citenamefont {Lukin}}]{LeePRL13}%
  \BibitemOpen
  \bibinfo {author} {T.~E. Lee}, \bibinfo {author} {S.~Gopalakrishnan}\ and\
  \bibinfo {author} {M.~D. Lukin},\ \emph {\bibinfo {title} {Unconventional
  Magnetism via Optical Pumping of Interacting Spin Systems}},\ \href
  {https://doi.org/10.1103/PhysRevLett.110.257204} {\bibfield  {journal}
  {\bibinfo  {journal} {Phys. Rev. Lett.}\ }\textbf {\bibinfo {volume} {110}},\
  \bibinfo {pages} {257204} (\bibinfo {year} {2013})}\BibitemShut {NoStop}%
\bibitem [{\citenamefont {Jin}\ \emph {et~al.}(2016)\citenamefont {Jin},
  \citenamefont {Biella}, \citenamefont {Viyuela}, \citenamefont {Mazza},
  \citenamefont {Keeling}, \citenamefont {Fazio},\ and\ \citenamefont
  {Rossini}}]{JinPRX16}%
  \BibitemOpen
  \bibinfo {author} {J.~Jin}, \bibinfo {author} {A.~Biella}, \bibinfo {author}
  {O.~Viyuela}, \bibinfo {author} {L.~Mazza}, \bibinfo {author} {J.~Keeling},
  \bibinfo {author} {R.~Fazio}\ and\ \bibinfo {author} {D.~Rossini},\ \emph
  {\bibinfo {title} {Cluster Mean-Field Approach to the Steady-State Phase
  Diagram of Dissipative Spin Systems}},\ \href
  {http://doi.org/10.1103/PhysRevX.6.031011} {\bibfield  {journal} {\bibinfo
  {journal} {Phys. Rev. X}\ }\textbf {\bibinfo {volume} {6}},\ \bibinfo {pages}
  {031011} (\bibinfo {year} {2016})}\BibitemShut {NoStop}%
\bibitem [{\citenamefont {Verstraelen}\ \emph {et~al.}(2023)\citenamefont
  {Verstraelen}, \citenamefont {Huybrechts}, \citenamefont {Roscilde},\ and\
  \citenamefont {Wouters}}]{Verstraelenetal2023}%
  \BibitemOpen
  \bibinfo {author} {W.~Verstraelen}, \bibinfo {author} {D.~Huybrechts},
  \bibinfo {author} {T.~Roscilde}\ and\ \bibinfo {author} {M.~Wouters},\ \emph
  {\bibinfo {title} {Quantum and Classical Correlations in Open Quantum Spin
  Lattices via Truncated-Cumulant Trajectories}},\ \href
  {https://doi.org/10.1103/PRXQuantum.4.030304} {\bibfield  {journal} {\bibinfo
   {journal} {PRX Quantum}\ }\textbf {\bibinfo {volume} {4}},\ \bibinfo {pages}
  {030304} (\bibinfo {year} {2023})}\BibitemShut {NoStop}%
\bibitem [{\citenamefont {Overbeck}\ \emph {et~al.}(2017)\citenamefont
  {Overbeck}, \citenamefont {Maghrebi}, \citenamefont {Gorshkov},\ and\
  \citenamefont {Weimer}}]{OverbeckPRA17}%
  \BibitemOpen
  \bibinfo {author} {V.~R. Overbeck}, \bibinfo {author} {M.~F. Maghrebi},
  \bibinfo {author} {A.~V. Gorshkov}\ and\ \bibinfo {author} {H.~Weimer},\
  \emph {\bibinfo {title} {Multicritical behavior in dissipative {I}sing
  models}},\ \href {\doibase 10.1103/PhysRevA.95.042133} {\bibfield  {journal}
  {\bibinfo  {journal} {Phys. Rev. A}\ }\textbf {\bibinfo {volume} {95}},\
  \bibinfo {pages} {042133} (\bibinfo {year} {2017})}\BibitemShut {NoStop}%
\bibitem [{\citenamefont {Finazzi}\ \emph {et~al.}(2015)\citenamefont
  {Finazzi}, \citenamefont {Le~Boit\'e}, \citenamefont {Storme}, \citenamefont
  {Baksic},\ and\ \citenamefont {Ciuti}}]{FinazziPRL15}%
  \BibitemOpen
  \bibinfo {author} {S.~Finazzi}, \bibinfo {author} {A.~Le~Boit\'e}, \bibinfo
  {author} {F.~Storme}, \bibinfo {author} {A.~Baksic}\ and\ \bibinfo {author}
  {C.~Ciuti},\ \emph {\bibinfo {title} {Corner-Space Renormalization Method for
  Driven-Dissipative Two-Dimensional Correlated Systems}},\ \href
  {https://doi.org/10.1103/PhysRevLett.115.080604} {\bibfield  {journal}
  {\bibinfo  {journal} {Phys. Rev. Lett.}\ }\textbf {\bibinfo {volume} {115}},\
  \bibinfo {pages} {080604} (\bibinfo {year} {2015})}\BibitemShut {NoStop}%
\bibitem [{\citenamefont {Weimer}(2015)}]{WeimerPRL2015}%
  \BibitemOpen
  \bibinfo {author} {H.~Weimer},\ \emph {\bibinfo {title} {Variational
  Principle for Steady States of Dissipative Quantum Many-Body Systems}},\
  \href {\doibase 10.1103/PhysRevLett.114.040402} {\bibfield  {journal}
  {\bibinfo  {journal} {Phys. Rev. Lett.}\ }\textbf {\bibinfo {volume} {114}},\
  \bibinfo {pages} {040402} (\bibinfo {year} {2015})}\BibitemShut {NoStop}%
\bibitem [{\citenamefont {Sieberer}\ \emph {et~al.}(2016)\citenamefont
  {Sieberer}, \citenamefont {Buchhold},\ and\ \citenamefont
  {Diehl}}]{Sieberer_2016}%
  \BibitemOpen
  \bibinfo {author} {L.~M. Sieberer}, \bibinfo {author} {M.~Buchhold}\ and\
  \bibinfo {author} {S.~Diehl},\ \emph {\bibinfo {title} {Keldysh field theory
  for driven open quantum systems}},\ \href
  {https://doi.org/10.1088/0034-4885/79/9/096001} {\bibfield  {journal}
  {\bibinfo  {journal} {Reports on Progress in Physics}\ }\textbf {\bibinfo
  {volume} {79}},\ \bibinfo {pages} {096001} (\bibinfo {year}
  {2016})}\BibitemShut {NoStop}%
\bibitem [{\citenamefont {Shammah}\ \emph {et~al.}(2018)\citenamefont
  {Shammah}, \citenamefont {Ahmed}, \citenamefont {Lambert}, \citenamefont
  {De~Liberato},\ and\ \citenamefont {Nori}}]{ShammahPRA98}%
  \BibitemOpen
  \bibinfo {author} {N.~Shammah}, \bibinfo {author} {S.~Ahmed}, \bibinfo
  {author} {N.~Lambert}, \bibinfo {author} {S.~De~Liberato}\ and\ \bibinfo
  {author} {F.~Nori},\ \emph {\bibinfo {title} {Open quantum systems with local
  and collective incoherent processes: Efficient numerical simulations using
  permutational invariance}},\ \href {\doibase 10.1103/PhysRevA.98.063815}
  {\bibfield  {journal} {\bibinfo  {journal} {Phys. Rev. A}\ }\textbf {\bibinfo
  {volume} {98}},\ \bibinfo {pages} {063815} (\bibinfo {year}
  {2018})}\BibitemShut {NoStop}%
\bibitem [{\citenamefont {Vicentini}\ \emph {et~al.}(2019)\citenamefont
  {Vicentini}, \citenamefont {Biella}, \citenamefont {Regnault},\ and\
  \citenamefont {Ciuti}}]{VicentiniPRL19}%
  \BibitemOpen
  \bibinfo {author} {F.~Vicentini}, \bibinfo {author} {A.~Biella}, \bibinfo
  {author} {N.~Regnault}\ and\ \bibinfo {author} {C.~Ciuti},\ \emph {\bibinfo
  {title} {Variational Neural-Network Ansatz for Steady States in Open Quantum
  Systems}},\ \href {\doibase 10.1103/PhysRevLett.122.250503} {\bibfield
  {journal} {\bibinfo  {journal} {Phys. Rev. Lett.}\ }\textbf {\bibinfo
  {volume} {122}},\ \bibinfo {pages} {250503} (\bibinfo {year}
  {2019})}\BibitemShut {NoStop}%
\bibitem [{\citenamefont {Nagy}\ and\ \citenamefont
  {Savona}(2019)}]{NagyPRL19}%
  \BibitemOpen
  \bibinfo {author} {A.~Nagy}\ and\ \bibinfo {author} {V.~Savona},\ \emph
  {\bibinfo {title} {Variational Quantum Monte Carlo Method with a
  Neural-Network Ansatz for Open Quantum Systems}},\ \href
  {https://doi.org/10.1103/PhysRevLett.122.250501} {\bibfield  {journal}
  {\bibinfo  {journal} {Phys. Rev. Lett.}\ }\textbf {\bibinfo {volume} {122}},\
  \bibinfo {pages} {250501} (\bibinfo {year} {2019})}\BibitemShut {NoStop}%
\bibitem [{\citenamefont {Hartmann}\ and\ \citenamefont
  {Carleo}(2019)}]{HartmannPRL19}%
  \BibitemOpen
  \bibinfo {author} {M.~J. Hartmann}\ and\ \bibinfo {author} {G.~Carleo},\
  \emph {\bibinfo {title} {Neural-Network Approach to Dissipative Quantum
  Many-Body Dynamics}},\ \href {\doibase 10.1103/PhysRevLett.122.250502}
  {\bibfield  {journal} {\bibinfo  {journal} {Phys. Rev. Lett.}\ }\textbf
  {\bibinfo {volume} {122}},\ \bibinfo {pages} {250502} (\bibinfo {year}
  {2019})}\BibitemShut {NoStop}%
\bibitem [{\citenamefont {Weimer}\ \emph {et~al.}(2021)\citenamefont {Weimer},
  \citenamefont {Kshetrimayum},\ and\ \citenamefont {Or\'us}}]{Weimeretal2021}%
  \BibitemOpen
  \bibinfo {author} {H.~Weimer}, \bibinfo {author} {A.~Kshetrimayum}\ and\
  \bibinfo {author} {R.~Or\'us},\ \emph {\bibinfo {title} {Simulation methods
  for open quantum many-body systems}},\ \href
  {https://doi.org/10.1103/RevModPhys.93.015008} {\bibfield  {journal}
  {\bibinfo  {journal} {Rev. Mod. Phys.}\ }\textbf {\bibinfo {volume} {93}},\
  \bibinfo {pages} {015008} (\bibinfo {year} {2021})}\BibitemShut {NoStop}%
\bibitem [{\citenamefont {Deuar}\ \emph {et~al.}(2021)\citenamefont {Deuar},
  \citenamefont {Ferrier}, \citenamefont {Matuszewski}, \citenamefont {Orso},\
  and\ \citenamefont {Szymanska}}]{DeuarPRXQ21}%
  \BibitemOpen
  \bibinfo {author} {P.~Deuar}, \bibinfo {author} {A.~Ferrier}, \bibinfo
  {author} {M.~Matuszewski}, \bibinfo {author} {G.~Orso}\ and\ \bibinfo
  {author} {M.~H. Szymanska},\ \emph {\bibinfo {title} {Fully Quantum Scalable
  Description of Driven-Dissipative Lattice Models}},\ \href
  {https://doi.org/10.1103/PRXQuantum.2.010319} {\bibfield  {journal} {\bibinfo
   {journal} {PRX Quantum}\ }\textbf {\bibinfo {volume} {2}},\ \bibinfo {pages}
  {010319} (\bibinfo {year} {2021})}\BibitemShut {NoStop}%
\bibitem [{\citenamefont {Mink}\ \emph {et~al.}(2022)\citenamefont {Mink},
  \citenamefont {Petrosyan},\ and\ \citenamefont
  {Fleischhauer}}]{mink2022hybrid}%
  \BibitemOpen
  \bibinfo {author} {C.~D. Mink}, \bibinfo {author} {D.~Petrosyan}\ and\
  \bibinfo {author} {M.~Fleischhauer},\ \emph {\bibinfo {title} {Hybrid
  discrete-continuous truncated Wigner approximation for driven, dissipative
  spin systems}},\ \href {https://doi.org/10.1103/PhysRevResearch.4.043136}
  {\bibfield  {journal} {\bibinfo  {journal} {Phys. Rev. Res.}\ }\textbf
  {\bibinfo {volume} {4}},\ \bibinfo {pages} {043136} (\bibinfo {year}
  {2022})}\BibitemShut {NoStop}%
\bibitem [{\citenamefont {Reitz}\ \emph {et~al.}(2022)\citenamefont {Reitz},
  \citenamefont {Sommer},\ and\ \citenamefont {Genes}}]{ReitzPRXQuantum2022}%
  \BibitemOpen
  \bibinfo {author} {M.~Reitz}, \bibinfo {author} {C.~Sommer}\ and\ \bibinfo
  {author} {C.~Genes},\ \emph {\bibinfo {title} {Cooperative Quantum Phenomena
  in Light-Matter Platforms}},\ \href {\doibase 10.1103/PRXQuantum.3.010201}
  {\bibfield  {journal} {\bibinfo  {journal} {PRX Quantum}\ }\textbf {\bibinfo
  {volume} {3}},\ \bibinfo {pages} {010201} (\bibinfo {year}
  {2022})}\BibitemShut {NoStop}%
\bibitem [{\citenamefont {Pezz\`e}\ \emph {et~al.}(2018)\citenamefont
  {Pezz\`e}, \citenamefont {Smerzi}, \citenamefont {Oberthaler}, \citenamefont
  {Schmied},\ and\ \citenamefont {Treutlein}}]{PezzeRMP2018}%
  \BibitemOpen
  \bibinfo {author} {L.~Pezz\`e}, \bibinfo {author} {A.~Smerzi}, \bibinfo
  {author} {M.~K. Oberthaler}, \bibinfo {author} {R.~Schmied}\ and\ \bibinfo
  {author} {P.~Treutlein},\ \emph {\bibinfo {title} {Quantum metrology with
  nonclassical states of atomic ensembles}},\ \href
  {https://doi.org/10.1103/RevModPhys.90.035005} {\bibfield  {journal}
  {\bibinfo  {journal} {Rev. Mod. Phys.}\ }\textbf {\bibinfo {volume} {90}},\
  \bibinfo {pages} {035005} (\bibinfo {year} {2018})}\BibitemShut {NoStop}%
\bibitem [{\citenamefont {Cong}\ \emph {et~al.}(2016)\citenamefont {Cong},
  \citenamefont {Zhang}, \citenamefont {Wang}, \citenamefont {Noe},
  \citenamefont {Belyanin},\ and\ \citenamefont {Kono}}]{Cong16}%
  \BibitemOpen
  \bibinfo {author} {K.~Cong}, \bibinfo {author} {Q.~Zhang}, \bibinfo {author}
  {Y.~Wang}, \bibinfo {author} {G.~T. Noe}, \bibinfo {author} {A.~Belyanin}\
  and\ \bibinfo {author} {J.~Kono},\ \emph {\bibinfo {title} {Dicke
  superradiance in solids}},\ \href {\doibase 10.1364/JOSAB.33.000C80}
  {\bibfield  {journal} {\bibinfo  {journal} {J. Opt. Soc. Am. B}\ }\textbf
  {\bibinfo {volume} {33}},\ \bibinfo {pages} {C80} (\bibinfo {year}
  {2016})}\BibitemShut {NoStop}%
\bibitem [{\citenamefont {Ferioli}\ \emph {et~al.}(2021)\citenamefont
  {Ferioli}, \citenamefont {Glicenstein}, \citenamefont {Robicheaux},
  \citenamefont {Sutherland}, \citenamefont {Browaeys},\ and\ \citenamefont
  {Ferrier-Barbut}}]{Ferioli21}%
  \BibitemOpen
  \bibinfo {author} {G.~Ferioli}, \bibinfo {author} {A.~Glicenstein}, \bibinfo
  {author} {F.~Robicheaux}, \bibinfo {author} {R.~T. Sutherland}, \bibinfo
  {author} {A.~Browaeys}\ and\ \bibinfo {author} {I.~Ferrier-Barbut},\ \emph
  {\bibinfo {title} {Laser-Driven Superradiant Ensembles of Two-Level Atoms
  near Dicke Regime}},\ \href {\doibase 10.1103/PhysRevLett.127.243602}
  {\bibfield  {journal} {\bibinfo  {journal} {Phys. Rev. Lett.}\ }\textbf
  {\bibinfo {volume} {127}},\ \bibinfo {pages} {243602} (\bibinfo {year}
  {2021})}\BibitemShut {NoStop}%
\bibitem [{\citenamefont {Breuer}\ and\ \citenamefont
  {Petruccione}(2007)}]{BreuerBookOpen}%
  \BibitemOpen
  \bibinfo {author} {H.~Breuer}\ and\ \bibinfo {author} {F.~Petruccione},\
  \href {https://doi.org/10.1093/acprof:oso/9780199213900.001.0001} {\emph
  {\bibinfo {title} {The Theory of Open Quantum Systems}}}\ (\bibinfo
  {publisher} {OUP Oxford},\ \bibinfo {year} {2007})\BibitemShut {NoStop}%
\bibitem [{\citenamefont {Benatti}\ \emph {et~al.}(2011)\citenamefont
  {Benatti}, \citenamefont {Nagy},\ and\ \citenamefont
  {Narnhofer}}]{Benatti_JPA2011}%
  \BibitemOpen
  \bibinfo {author} {F.~Benatti}, \bibinfo {author} {A.~Nagy}\ and\ \bibinfo
  {author} {H.~Narnhofer},\ \emph {\bibinfo {title} {Asymptotic entanglement
  and Lindblad dynamics: a perturbative approach}},\ \href
  {https://doi.org/10.1088/1751-8113/44/15/155303} {\bibfield  {journal}
  {\bibinfo  {journal} {Journal of Physics A: Mathematical and Theoretical}\
  }\textbf {\bibinfo {volume} {44}},\ \bibinfo {pages} {155303} (\bibinfo
  {year} {2011})}\BibitemShut {NoStop}%
\bibitem [{\citenamefont {Li}\ \emph {et~al.}(2014)\citenamefont {Li},
  \citenamefont {Petruccione},\ and\ \citenamefont {Koch}}]{Li14}%
  \BibitemOpen
  \bibinfo {author} {A.~C.~Y. Li}, \bibinfo {author} {F.~Petruccione}\ and\
  \bibinfo {author} {J.~Koch},\ \emph {\bibinfo {title} {Perturbative approach
  to Markovian open quantum systems}},\ \href {\doibase 10.1038/srep04887}
  {\bibfield  {journal} {\bibinfo  {journal} {Scientific Reports}\ }\textbf
  {\bibinfo {volume} {4}},\ \bibinfo {pages} {4887} (\bibinfo {year}
  {2014})}\BibitemShut {NoStop}%
\bibitem [{\citenamefont {Li}\ \emph {et~al.}(2016)\citenamefont {Li},
  \citenamefont {Petruccione},\ and\ \citenamefont {Koch}}]{LiPRX16}%
  \BibitemOpen
  \bibinfo {author} {A.~C.~Y. Li}, \bibinfo {author} {F.~Petruccione}\ and\
  \bibinfo {author} {J.~Koch},\ \emph {\bibinfo {title} {Resummation for
  Nonequilibrium Perturbation Theory and Application to Open Quantum
  Lattices}},\ \href {\doibase 10.1103/PhysRevX.6.021037} {\bibfield  {journal}
  {\bibinfo  {journal} {Phys. Rev. X}\ }\textbf {\bibinfo {volume} {6}},\
  \bibinfo {pages} {021037} (\bibinfo {year} {2016})}\BibitemShut {NoStop}%
\bibitem [{\citenamefont {G{\'o}mez}\ \emph {et~al.}(2018)\citenamefont
  {G{\'o}mez}, \citenamefont {Casta{\~n}o-Yepes},\ and\ \citenamefont
  {Thirumuruganandham}}]{GomezRiP18}%
  \BibitemOpen
  \bibinfo {author} {E.~A. G{\'o}mez}, \bibinfo {author} {J.~D.
  Casta{\~n}o-Yepes}\ and\ \bibinfo {author} {S.~P. Thirumuruganandham},\ \emph
  {\bibinfo {title} {Perturbation theory for open quantum systems at the steady
  state}},\ \href {https://doi.org/10.1016/j.rinp.2018.06.038} {\bibfield
  {journal} {\bibinfo  {journal} {Results in Physics}\ }\textbf {\bibinfo
  {volume} {10}},\ \bibinfo {pages} {353} (\bibinfo {year} {2018})}\BibitemShut
  {NoStop}%
\bibitem [{\citenamefont {Lenar\ifmmode \check{c}\else
  \v{c}\fi{}i\ifmmode~\check{c}\else \v{c}\fi{}}\ \emph
  {et~al.}(2018)\citenamefont {Lenar\ifmmode \check{c}\else
  \v{c}\fi{}i\ifmmode~\check{c}\else \v{c}\fi{}}, \citenamefont {Lange},\ and\
  \citenamefont {Rosch}}]{LenarcicPRB18}%
  \BibitemOpen
  \bibinfo {author} {Z.~Lenar\ifmmode \check{c}\else
  \v{c}\fi{}i\ifmmode~\check{c}\else \v{c}\fi{}}, \bibinfo {author} {F.~Lange}\
  and\ \bibinfo {author} {A.~Rosch},\ \emph {\bibinfo {title} {Perturbative
  approach to weakly driven many-particle systems in the presence of
  approximate conservation laws}},\ \href {\doibase 10.1103/PhysRevB.97.024302}
  {\bibfield  {journal} {\bibinfo  {journal} {Phys. Rev. B}\ }\textbf {\bibinfo
  {volume} {97}},\ \bibinfo {pages} {024302} (\bibinfo {year}
  {2018})}\BibitemShut {NoStop}%
\bibitem [{\citenamefont {Kato}(1995)}]{KatoBOOK}%
  \BibitemOpen
  \bibinfo {author} {T.~Kato},\ \href
  {https://doi.org/10.1007/978-3-642-66282-9} {\emph {\bibinfo {title}
  {Perturbation theory for linear operators}}},\ Classics in Mathematics\
  (\bibinfo  {publisher} {Springer},\ \bibinfo {year} {1995})\BibitemShut
  {NoStop}%
\bibitem [{\citenamefont {Kessler}(2012)}]{Kessler2012}%
  \BibitemOpen
  \bibinfo {author} {E.~M. Kessler},\ \emph {\bibinfo {title} {Generalized
  Schrieffer-Wolff formalism for dissipative systems}},\ \href
  {https://doi.org/10.1103/PhysRevA.86.012126} {\bibfield  {journal} {\bibinfo
  {journal} {Phys. Rev. A}\ }\textbf {\bibinfo {volume} {86}},\ \bibinfo
  {pages} {012126} (\bibinfo {year} {2012})}\BibitemShut {NoStop}%
\bibitem [{\citenamefont {Landau}\ and\ \citenamefont
  {Lifshitz}(1981)}]{Landau_BOOK_Quantum}%
  \BibitemOpen
  \bibinfo {author} {L.~Landau}\ and\ \bibinfo {author} {E.~Lifshitz},\
  \href@noop {} {\emph {\bibinfo {title} {Quantum Mechanics: Non-Relativistic
  Theory}}},\ \bibinfo {series} {Course of Theoretical Physics}, Vol.~\bibinfo
  {volume} {3}\ (\bibinfo  {publisher} {Elsevier Science},\ \bibinfo {year}
  {1981})\BibitemShut {NoStop}%
\bibitem [{\citenamefont {Wineland}\ \emph {et~al.}(1994)\citenamefont
  {Wineland}, \citenamefont {Bollinger}, \citenamefont {Itano},\ and\
  \citenamefont {Heinzen}}]{WinelandPRA1994}%
  \BibitemOpen
  \bibinfo {author} {D.~J. Wineland}, \bibinfo {author} {J.~J. Bollinger},
  \bibinfo {author} {W.~M. Itano}\ and\ \bibinfo {author} {D.~J. Heinzen},\
  \emph {\bibinfo {title} {Squeezed atomic states and projection noise in
  spectroscopy}},\ \href {\doibase 10.1103/PhysRevA.50.67} {\bibfield
  {journal} {\bibinfo  {journal} {Phys. Rev. A}\ }\textbf {\bibinfo {volume}
  {50}},\ \bibinfo {pages} {67} (\bibinfo {year} {1994})}\BibitemShut {NoStop}%
\bibitem [{\citenamefont {S{\o}rensen}\ \emph {et~al.}(2001)\citenamefont
  {S{\o}rensen}, \citenamefont {Duan}, \citenamefont {Cirac},\ and\
  \citenamefont {Zoller}}]{Sorensen2001}%
  \BibitemOpen
  \bibinfo {author} {A.~S{\o}rensen}, \bibinfo {author} {L.-M. Duan}, \bibinfo
  {author} {J.~I. Cirac}\ and\ \bibinfo {author} {P.~Zoller},\ \emph {\bibinfo
  {title} {Many-particle entanglement with Bose--Einstein condensates}},\ \href
  {\doibase 10.1038/35051038} {\bibfield  {journal} {\bibinfo  {journal}
  {Nature}\ }\textbf {\bibinfo {volume} {409}},\ \bibinfo {pages} {63}
  (\bibinfo {year} {2001})}\BibitemShut {NoStop}%
\bibitem [{\citenamefont {Braunstein}\ and\ \citenamefont
  {Caves}(1994)}]{Braunstein94}%
  \BibitemOpen
  \bibinfo {author} {S.~L. Braunstein}\ and\ \bibinfo {author} {C.~M. Caves},\
  \emph {\bibinfo {title} {Statistical distance and the geometry of quantum
  states}},\ \href {\doibase 10.1103/PhysRevLett.72.3439} {\bibfield  {journal}
  {\bibinfo  {journal} {Phys. Rev. Lett.}\ }\textbf {\bibinfo {volume} {72}},\
  \bibinfo {pages} {3439} (\bibinfo {year} {1994})}\BibitemShut {NoStop}%
\bibitem [{\citenamefont {Roscilde}\ \emph {et~al.}(2021)\citenamefont
  {Roscilde}, \citenamefont {Mezzacapo},\ and\ \citenamefont
  {Comparin}}]{Roscildeetal2022}%
  \BibitemOpen
  \bibinfo {author} {T.~Roscilde}, \bibinfo {author} {F.~Mezzacapo}\ and\
  \bibinfo {author} {T.~Comparin},\ \emph {\bibinfo {title} {Spin squeezing
  from bilinear spin-spin interactions: Two simple theorems}},\ \href
  {https://doi.org/10.1103/PhysRevA.104.L040601} {\bibfield  {journal}
  {\bibinfo  {journal} {Phys. Rev. A}\ }\textbf {\bibinfo {volume} {104}},\
  \bibinfo {pages} {L040601} (\bibinfo {year} {2021})}\BibitemShut {NoStop}%
\bibitem [{\citenamefont {Huybrechts}\ \emph {et~al.}(2020)\citenamefont
  {Huybrechts}, \citenamefont {Minganti}, \citenamefont {Nori}, \citenamefont
  {Wouters},\ and\ \citenamefont {Shammah}}]{HuybrechtsPRB20}%
  \BibitemOpen
  \bibinfo {author} {D.~Huybrechts}, \bibinfo {author} {F.~Minganti}, \bibinfo
  {author} {F.~Nori}, \bibinfo {author} {M.~Wouters}\ and\ \bibinfo {author}
  {N.~Shammah},\ \emph {\bibinfo {title} {Validity of mean-field theory in a
  dissipative critical system: Liouvillian gap, $\mathbb{PT}$-symmetric
  antigap, and permutational symmetry in the $XYZ$ model}},\ \href
  {https://doi.org/10.1103/PhysRevB.101.214302} {\bibfield  {journal} {\bibinfo
   {journal} {Phys. Rev. B}\ }\textbf {\bibinfo {volume} {101}},\ \bibinfo
  {pages} {214302} (\bibinfo {year} {2020})}\BibitemShut {NoStop}%
\bibitem [{\citenamefont {Johansson}\ \emph {et~al.}(2012)\citenamefont
  {Johansson}, \citenamefont {Nation},\ and\ \citenamefont {Nori}}]{qutip1}%
  \BibitemOpen
  \bibinfo {author} {J.~Johansson}, \bibinfo {author} {P.~Nation}\ and\
  \bibinfo {author} {F.~Nori},\ \emph {\bibinfo {title} {{QuTiP}: An
  open-source Python framework for the dynamics of open quantum systems}},\
  \href {\doibase 10.1016/j.cpc.2012.02.021} {\bibfield  {journal} {\bibinfo
  {journal} {Computer Physics Communications}\ }\textbf {\bibinfo {volume}
  {183}},\ \bibinfo {pages} {1760} (\bibinfo {year} {2012})}\BibitemShut
  {NoStop}%
\bibitem [{\citenamefont {Johansson}\ \emph {et~al.}(2013)\citenamefont
  {Johansson}, \citenamefont {Nation},\ and\ \citenamefont {Nori}}]{qutip2}%
  \BibitemOpen
  \bibinfo {author} {J.~Johansson}, \bibinfo {author} {P.~Nation}\ and\
  \bibinfo {author} {F.~Nori},\ \emph {\bibinfo {title} {{QuTiP} 2: A Python
  framework for the dynamics of open quantum systems}},\ \href
  {https://doi.org/10.1016/j.cpc.2012.11.019} {\bibfield  {journal} {\bibinfo
  {journal} {Computer Physics Communications}\ }\textbf {\bibinfo {volume}
  {184}},\ \bibinfo {pages} {1234} (\bibinfo {year} {2013})}\BibitemShut
  {NoStop}%
\bibitem [{\citenamefont {Rota}\ \emph {et~al.}(2017)\citenamefont {Rota},
  \citenamefont {Storme}, \citenamefont {Bartolo}, \citenamefont {Fazio},\ and\
  \citenamefont {Ciuti}}]{RotaPRB17}%
  \BibitemOpen
  \bibinfo {author} {R.~Rota}, \bibinfo {author} {F.~Storme}, \bibinfo {author}
  {N.~Bartolo}, \bibinfo {author} {R.~Fazio}\ and\ \bibinfo {author}
  {C.~Ciuti},\ \emph {\bibinfo {title} {Critical behavior of dissipative
  two-dimensional spin lattices}},\ \href
  {https://doi.org/10.1103/PhysRevB.95.134431} {\bibfield  {journal} {\bibinfo
  {journal} {Phys. Rev. B}\ }\textbf {\bibinfo {volume} {95}},\ \bibinfo
  {pages} {134431} (\bibinfo {year} {2017})}\BibitemShut {NoStop}%
\bibitem [{\citenamefont {Rota}\ \emph {et~al.}(2018)\citenamefont {Rota},
  \citenamefont {Minganti}, \citenamefont {Biella},\ and\ \citenamefont
  {Ciuti}}]{RotaNJP18}%
  \BibitemOpen
  \bibinfo {author} {R.~Rota}, \bibinfo {author} {F.~Minganti}, \bibinfo
  {author} {A.~Biella}\ and\ \bibinfo {author} {C.~Ciuti},\ \emph {\bibinfo
  {title} {Dynamical properties of dissipative {X}{Y}{Z} {H}eisenberg
  lattices}},\ \href {http://doi.org/10.1088/1367-2630/aab703} {\bibfield
  {journal} {\bibinfo  {journal} {New Journal of Physics}\ }\textbf {\bibinfo
  {volume} {20}},\ \bibinfo {pages} {045003} (\bibinfo {year}
  {2018})}\BibitemShut {NoStop}%
\bibitem [{\citenamefont {Huybrechts}\ and\ \citenamefont
  {Wouters}(2019)}]{HuybrechtsPRA19}%
  \BibitemOpen
  \bibinfo {author} {D.~Huybrechts}\ and\ \bibinfo {author} {M.~Wouters},\
  \emph {\bibinfo {title} {Cluster methods for the description of a
  driven-dissipative spin model}},\ \href {\doibase 10.1103/PhysRevA.99.043841}
  {\bibfield  {journal} {\bibinfo  {journal} {Phys. Rev. A}\ }\textbf {\bibinfo
  {volume} {99}},\ \bibinfo {pages} {043841} (\bibinfo {year}
  {2019})}\BibitemShut {NoStop}%
\bibitem [{\citenamefont {Huybrechts}\ and\ \citenamefont
  {Roscilde}(2026)}]{Huybrechts_in_preparation}%
  \BibitemOpen
  \bibinfo {author} {D.~Huybrechts}\ and\ \bibinfo {author} {T.~Roscilde},\
  \href@noop {} {\bibfield  {journal} {\bibinfo  {journal} {In Preparation}\ }
  (\bibinfo {year} {2026})}\BibitemShut {NoStop}%
\bibitem [{\citenamefont {Carollo}\ and\ \citenamefont
  {Lesanovsky}(2021)}]{CarolloPRL21}%
  \BibitemOpen
  \bibinfo {author} {F.~Carollo}\ and\ \bibinfo {author} {I.~Lesanovsky},\
  \emph {\bibinfo {title} {Exactness of Mean-Field Equations for Open Dicke
  Models with an Application to Pattern Retrieval Dynamics}},\ \href
  {https://doi.org/10.1103/PhysRevLett.126.230601} {\bibfield  {journal}
  {\bibinfo  {journal} {Phys. Rev. Lett.}\ }\textbf {\bibinfo {volume} {126}},\
  \bibinfo {pages} {230601} (\bibinfo {year} {2021})}\BibitemShut {NoStop}%
\bibitem [{\citenamefont {Narducci}\ \emph {et~al.}(1978)\citenamefont
  {Narducci}, \citenamefont {Feng}, \citenamefont {Gilmore},\ and\
  \citenamefont {Agarwal}}]{Narduccietal1978}%
  \BibitemOpen
  \bibinfo {author} {L.~M. Narducci}, \bibinfo {author} {D.~H. Feng}, \bibinfo
  {author} {R.~Gilmore}\ and\ \bibinfo {author} {G.~S. Agarwal},\ \emph
  {\bibinfo {title} {Transient and steady-state behavior of collective atomic
  systems driven by a classical field}},\ \href
  {https://doi.org/10.1103/PhysRevA.18.1571} {\bibfield  {journal} {\bibinfo
  {journal} {Phys. Rev. A}\ }\textbf {\bibinfo {volume} {18}},\ \bibinfo
  {pages} {1571} (\bibinfo {year} {1978})}\BibitemShut {NoStop}%
\bibitem [{\citenamefont {Hannukainen}\ and\ \citenamefont
  {Larson}(2018)}]{HannukainenL2018}%
  \BibitemOpen
  \bibinfo {author} {J.~Hannukainen}\ and\ \bibinfo {author} {J.~Larson},\
  \emph {\bibinfo {title} {Dissipation-driven quantum phase transitions and
  symmetry breaking}},\ \href {\doibase 10.1103/PhysRevA.98.042113} {\bibfield
  {journal} {\bibinfo  {journal} {Phys. Rev. A}\ }\textbf {\bibinfo {volume}
  {98}},\ \bibinfo {pages} {042113} (\bibinfo {year} {2018})}\BibitemShut
  {NoStop}%
\end{thebibliography}
